\documentclass[aps,pre,twocolumn,superscriptaddress,10pt]{revtex4}

\usepackage[dvips]{graphics}
\usepackage{graphicx}
\usepackage{amsfonts}
\usepackage{color}
\usepackage{amssymb}
\usepackage{amsmath}

\begin{document}

\title{\textbf{
Faraday waves in Bose-Einstein condensates}}

\author{Alexandru I. Nicolin}
\affiliation{Niels Bohr Institute, Blegdamsvej 17,
Copenhagen, DK-2100, Denmark}

\author{R. Carretero-Gonz\'alez}
\affiliation{Nonlinear Dynamical Systems Group%
\footnote{URL: {\tt http://nlds.sdsu.edu}},
Department of Mathematics and Statistics,
and Computational Science Research Center,
San Diego State University, San Diego CA, 92182-7720, USA}

\author{P.G.\ Kevrekidis}
\affiliation{Department of Mathematics and Statistics,
University of Massachusetts, Amherst MA 01003-4515, USA}

\date{Submitted to {\em Phys.~Rev.~A}, October 2007.}

\begin{abstract}
Motivated by recent experiments on Faraday waves in Bose-Einstein
condensates we investigate both analytically and numerically the
dynamics of cigar-shaped Bose-condensed gases subject to periodic
modulation of the strength of the transverse confinement. We offer
a fully analytical explanation of the observed parametric
resonance, based on a Mathieu-type analysis of the non-polynomial
Schr{\"o}dinger equation. The theoretical prediction for the
pattern periodicity versus the driving frequency is directly
compared with the experimental data, yielding good qualitative and
quantitative agreement between the two. These results are
corroborated by direct numerical simulations of both the
one-dimensional non-polynomial Schr{\"o}dinger equation and of the
fully three-dimensional Gross-Pitaevskii equation.
\end{abstract}

\maketitle

\section{Introduction}

Pattern formation in driven systems is an important direction of
current research that influences many fields ranging from
hydrodynamics to biophysics and from nonlinear optics to reaction
kinetics; see \cite{cross} for a comprehensive review of the
topic.

Some of the oldest and most well-known forms of such phenomena are
the so-called Faraday patterns, stemming from the classical
studies of Faraday in 1831 \cite{far}, who studied the behavior of
``groups of particles [placed] upon vibrating elastic surfaces" and
(in the appendix of his much-celebrated paper) the dynamics of
``fluids in contact with vibrating surfaces." Faraday's experiment
became a classical example of pattern formation, whereby the
uniform state loses its stability against spatially modulated
waveforms, whose dominant length-scale is determined by the
intrinsic properties of the system (such as the dominant
wavelength of the instability) and is only weakly dependent on
boundary or initial conditions.

In the past few years, there has been an increasing literature
about observing phenomena of the above type in driven superfluids.
Bose-Einstein condensates \cite{book1,book2} offer perhaps the
ideal playground for such experiments, since their experimental
tunability permits to create such parametric resonance phenomena
in a multiplicity of ways. One such way is by driving the magnetic
trap confining the system, as was proposed in Ref.~\cite{perez}.
In the same spirit of modulating the confinement of the system
to observe parametric resonances, the works of
Refs.~\cite{dalfovo1,dalfovo2}, motivated by the experiments
of Ref.~\cite{stoferle}, considered a periodic modulation of an optical
lattice potential; on the other hand, the later work of
Ref.~\cite{dalfovo3} focused
on modulating in time the confinement potential in a toroidal trap.
Another recent suggestion was to produce a parametric drive
by means of periodically modulating the scattering length
\cite{staliunas}, which is directly related to the prefactor of
the effective nonlinearity (due to the mean-field inter-particle
interactions) of the system. This can be achieved by means of
Feshbach resonances \cite{feshbach}.

These theoretical propositions motivated the very recent
experimental implementation of the Faraday waves in Bose-Einstein
condensates in the work of Ref.~\cite{engels}. The actual
realization of the spatially modulated patterns arose in a somewhat
different way than was proposed in the above studies, a way which
is very close, however, to the spirit of the theoretical
suggestion of Ref.~\cite{staliunas_second} (see also the more
recent consideration of a similar problem from a quantum point
of view in Ref.~\cite{kagan}).
In particular, in Ref.~\cite{engels}, an elongated cigar-shaped condensate was
used where the {\it transverse}, strong confinement directions
were periodically modulated in time, while the weaker longitudinal
direction confinement was time-independent. The parametric
excitation at the driving frequency was recognized as being
responsible for exciting oscillations at half the driving
frequency, which is the main resonance also observed in Faraday's
experiments. Subsequently, based on this insight and the
dispersion relation of longitudinal collective modes presented in
Refs.~\cite{eng13,eng14}, a relation was derived (and convincingly
compared to the experimental results in a quantitative manner) for
the resulting pattern periodicity versus the transverse driving
frequency.

The aim of the present paper is to provide a complete analysis of
the instability from first principles and to obtain
a fully analytical prediction that can be used
for a detailed comparison with the experimental results and
numerical results obtained from the 1D model reduction as
well as full 3D simulations at the mean-field level.

The principal feature which allows us to provide a detailed
quantitative analysis of the system is the fact that for
cigar-shaped condensates (such as the ones used in the experiment
of Ref.~\cite{engels}) there is a quantitatively accurate
description in the form of the non-polynomial Schr{\"o}dinger
equation (NPSE) derived in Ref.~\cite{npse} (provided that the
transverse direction stays close to its ground state, which is
approximately the case in the experiments of Ref.~\cite{engels}).
The remarkable feature of the NPSE is that in the
resulting partial differential equation (PDE) for the longitudinal
description of the condensate, the transverse frequency enters
explicitly and hence provides, in this setting, the explicit
parametric drive that will lead to the observed spatially
modulated patterns. To elucidate this feature, we will perform a
linear stability analysis of the spatially uniform states in this
effectively one-dimensional setting. This naturally leads to a
Mathieu equation. Then, one can use the
theory of the Mathieu equation to identify the most unstable mode
and the wavenumber (and associated wavelength) of its spatial
periodicity. This, in turn, allows us to obtain a fully analytical
expression for the pattern periodicity as a function of the
driving frequency that can be directly compared with the
experiment. We believe that this approach yields both qualitative
and quantitative insight on the experiment and on the nature of
the relevant phenomena.

Our presentation will be structured as follows. In Section II, we develop
our analytical approach and directly compare it with the experimental
findings. In Section III, we complement our theoretical findings
with numerical results simulating both
the full 3D experimental setting and comparing it with results of
its 1D analog, namely the NPSE equation. Finally, in Section IV, we
summarize our findings and present some interesting directions for
future work.

\section{Analytical Considerations}

Our starting point will be the standard mean-field model for Bose-Einstein
condensates (BEC), namely the three-dimensional Gross-Pitaevskii (GP)
equation \cite{book1,book2}
\begin{equation}
i\hbar\frac{\partial\psi}{\partial t}=\left[-\frac{\hbar^{2}}{2m}\nabla^{2}
+V({\bf r})+gN\left|\psi\right|^{2}\right]\psi,
\label{GPE}
\end{equation}
where $\psi$ denotes the BEC wavefunction (normalized to unity),
$N$ represents the number of atoms and $g=4 \pi \hbar^2 a_s/m$ is
proportional to the
scattering length $a_s$ of the interatomic interactions; the external
potential confining the atoms is given by:
\begin{eqnarray*}
V(r,z) & = &
\frac{1}{2}m\omega_{r}^{2}r^{2}+U(z),
\end{eqnarray*}
where $r^2=x^2+y^2$, $\omega_r$ is the transverse trapping frequency,
and $U(z)$, in the case of Ref.~\cite{engels}, represents a far weaker
parabolic, longitudinal, potential (with a magnetic
trap frequency less than 5\%
of the transverse frequency), which will therefore be neglected
for the purposes of the (main analytical portion of the) present work.

Following the approach of Ref.~\cite{npse},
the three-dimensional wave function can be decomposed
into a radial and a longitudinal part as
\begin{equation}
\psi({\bf r},t)=\phi(r,t;\sigma(z,t))f(z,t),\label{decomposition_var}
\end{equation}
where the radial component is taken as
\begin{equation}
\phi(r,t;\sigma(z,t))=
\frac{\exp\left[-r^{2}/(2\sigma^{2}(z,t))\right]}
{\sqrt{\pi}\sigma(z,t)}, \label{gaussian_ansatz_var}
\end{equation}
with a spatially and temporally variable width characterized by
$\sigma(z,t)$. It should be noted here that the radial profiles of high-density
cigar-shaped condensates (as the one in Ref. \cite{engels}) are
closer to the Thomas-Fermi regime than to the Gaussian one.
While this introduces an element of approximation to the calculation below,
the adjustable nature of the parameter $\sigma(z,t)$ (which is always
larger than the transverse oscillator length by a spatially dependent
factor that depends on the longitudinal wavefunction; see  Ref.~\cite{npse})
and the comparison that we will report below between our theory and the
physical experiment render this approximation a reasonable one for
the purposes of predicting the wavelength of the
resulting pattern. We note in passing that
it appears to be an interesting open problem
to perform a derivation similar to that of Ref.~\cite{npse}, under
the assumption of a transverse Thomas-Fermi wavefunction profile.

Employing the standard variational recipe of Ref.~\cite{npse},
and neglecting the longitudinal potential $U(z)$,
one obtains the following effective PDE, the so-called
non-polynomial Schr{\"o}dinger equation (NPSE),
describing the longitudinal wavefunction:
\begin{eqnarray}
i\hbar\frac{\partial f}{\partial t}
%& = &
=\left[-\frac{\hbar^{2}}{2m} \frac{\partial^{2}}{\partial z^{2}}
%\frac{gNm\omega_{r}}{2\pi\hbar}\frac{\left|f\right|^{2}}{\sqrt{1+2\, a_{s}N\left|f\right|^{2}}}\right.\nonumber \\
%&  & \left.
+\hbar\omega_{r}\frac{1+3\, a_{s}N\left|f\right|^{2}}{\sqrt{1+2\, a_{s}N\left|f\right|^{2}}}\right]f.
\label{salasnich_eq}
\end{eqnarray}
%where we have neglected the longitudinal confinement. Considering
In accordance with the experimental setup of Ref.~\cite{engels},
the transverse frequency is modulated by
\begin{equation}
\omega_{r}(t)=\omega_{r,0}\cdot(1+\epsilon\sin(\omega t)),
\end{equation}
where $\omega_{r,0}$ is the reference tranverse frequency, and $\epsilon$
and $\omega$ are, respectively, the amplitude and frequency of the modulation.
Then, the spatially homogeneous solution is given by
\begin{equation}
f_{0}(t)=A\exp\left[-ic\left(t-\epsilon\frac{\cos(\omega t)}{\omega}\right)\right],\label{homogeneous_sol_var}\end{equation}
where \begin{equation}
c=\frac{\omega_{r,0}}{\sqrt{1+2\, a_{s}NA^{2}}}\left(
%\frac{gNmA^{2}}{2\pi\hbar}+\hbar+\hbar
1+3\, a_{s}NA^{2}\right)\label{constant_var}\end{equation}
and $A$ is a positive constant. The numerical value of $A$ is computed from
the normalization
%
%\begin{equation}
$
\int\!\!\int\!\!\int |\psi(r,z,t)|^2 \,d{\bf r} = 1.
$
%
%
%\begin{equation}
%\int_{-L}^{L}\int_{-\infty}^{\infty}\int_{-\infty}^{\infty}
%|\psi(r,z,t)|^2 \,dx\,dy\,dz = 1.
%\label{norm_var}
%\end{equation}
%
Computing the integral one has that $A=\sqrt{1/2L}$, where we
assume that the condensate extends between $-L$ and $L$;
i.e., we are assuming here that the condensate is in a box
rather than in a very weak magnetic trap in the longitudinal
direction. The validity of this assumption for the present
phenomenology is verified both a priori (due to the very weak
nature of the longitudinal confinement in comparison to the
much stronger transverse confinement and its modulation)
and a posteriori (from the comparison with the experimental
results).

Faraday patterns appear in this context due to a modulational instability
along
the (longitudinal) $z$-axis. To examine the modulational stability of uniform
patterns in the $z$-direction, we use the ansatz
\begin{equation}
f(t)=f_{0}(t)\left[1+(u(t)+iv(t))\cos(kz)\right].
\label{perturb_var}\end{equation}
Inserting this ansatz into Eq.~(\ref{salasnich_eq}) and linearizing
the ensuing equations yields a Mathieu-type equation for
the perturbation:
\begin{equation}
\frac{d^{2}u}{d{\tau}^{2}}+\left(a(k,\omega)+b(k,\omega)\sin(2{\tau})\right)u=0,\label{Mathieu_var}\end{equation}
where $a$ and \textbf{$b$} are given by Eqs.~(\ref{a_var}) and
(\ref{b_var}) and $\tau={\omega}t/2$.

\begin{widetext}
\begin{eqnarray}
a(k,\omega)&=&\frac{k^{2}}{2 \hbar {\pi} m^{2}\omega^{2}}\frac{6
\pi a_{s}^{2}\hbar^{2}mN^{2}
\omega_{r,0}+a_{s}N2\hbar^{3}k^{2}\pi\sqrt{L^{2}+La_{s}N}
%+a_{s}gm^{2}N^{2}\omega_{r,0}
+2\hbar^{3}k^{2}\pi\sqrt{L^{4}+L^{3}a_{s}N}+2gLm^{2}N\omega_{r,0}}{a_{s}N\sqrt{L^{2}+La_{s}N}+\sqrt{L^{4}+L^{3}a_{s}N}},
\label{a_var}
\\[2.0ex]
b(k,\omega)&=&\frac{k^{2}\omega_{r,0}\epsilon N}{2\hbar\pi m\omega^{2}}\frac{2gmL
%+gma_{s}N
+ 6
a_{s}^{2}\hbar^{2}N\pi}{a_{s}N\sqrt{L^{2}+La_{s}N}+\sqrt{L^{4}+L^{3}a_{s}N}}.
\label{b_var}
\end{eqnarray}
\end{widetext}

As is commonly known, Mathieu equation exhibits
an intricate stability chart comprising tongues of both stable and
unstable solutions \cite{book_of_McLachlan}.
Due to the periodic potential a generic solution takes the form
$u(t)=e^{i\mu t}g(t)$,
where $g(t)$ has the same periodicity as $\sin(2 \tau)$ (according to
the Floquet-Bloch theorem),
%\cite{Ince}),
and $\mu$ is a complex exponent
taken as $\mu=\mu_{1}+i\mu_{2}$, where both $\mu_{1}$ and $\mu_{2}$
are real numbers. Determining the most unstable mode (which is the
one that is expected to be observed experimentally)
amounts to finding the $\omega(k)$ curve
corresponding to the critical exponent with the most negative imaginary
part. While this is usually a complicated task, it  can be shown that for
small positive values of $b$ it amounts to
\begin{equation}
a(k,\omega)\approx1.\label{final_eq_var}\end{equation}
This conclusive property can be seen both numerically and analytically.
\begin{figure}[tbhp!]
     \centering
           \includegraphics[width=8cm]{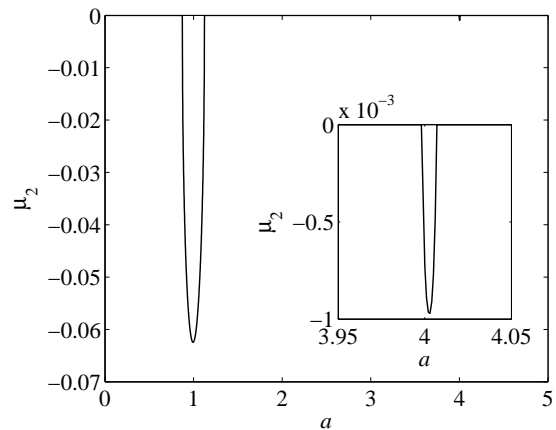}
      \caption{Imaginary part of the critical exponent $\mu$ as function of $a$ for $b=0.25$. Notice the main lobe centered around $a=1$.
      For the first lobe, the difference between the approximation (\ref{critical_exponent}) and the numerics is smaller than the thickness of the line.
      The second lobe, shown in the inset, is so much smaller than the first one that they can hardly be seen on the same scale.}
      \label{fig_ce}
\end{figure}

Investigating $\mu_{2}$ numerically \cite{mathematica} one finds
that (for small positive values of $b$) it consists of a set of
symmetrical lobes centered around $a=n^{2}$, where $n$ is a
positive integer, the one centered around $a=1$ being substantially
larger than the other ones (see Fig.~\ref{fig_ce}).
These lobes correspond to the unstable
regions. Inspecting these lobes, it is transparent that the most
unstable mode corresponds to $a \approx 1$.

One can also argue for the validity of Eq.~(\ref{final_eq_var})
by analytical means. There is a class of
partly-forgotten approximate formulas for $\mu$ as a function of
$a$ and $b$ stemming from celestial mechanics
(see Ref.~\cite{critical_exponent} for the main results). A convenient
formula for our purpose is

\begin{equation}
\mu=\frac{1}{\pi}\arccos\left[\cos\left(\pi\sqrt{a}\right)+\frac{\pi b^{2}\sin\left(\pi\sqrt{a}\right)}{16\sqrt{a}\left(a-1\right)}+\mathcal{O}\left(b^{4}\right)\right],\label{critical_exponent}\end{equation}
which describes accurately the first lobe of $\mu_{2}$ for small
values of $b$. In order for $\mu$ to have an imaginary part the
argument of $\arccos$ must be larger than one (in absolute value).
Naturally, identifying the most unstable mode amounts to finding the
extremum value of
$$
\cos\left(\pi\sqrt{a}\right)+\frac{\pi b^{2}\sin\left(\pi\sqrt{a}\right)}{16\sqrt{a}\left(a-1\right)}.
$$
To leading order in $b$ this corresponds to $a=1$.

Finally, numerically solving Eq.~(\ref{final_eq_var}), one readily
obtains $2 \pi/k$, which represents the spacing of adjacent maxima,
herein called $\cal{S}$, as a function of the driving frequency
$\omega$ of the transverse confinement $\omega_{r}$. Neglecting
the $k^4$ term in Eq.~(\ref{final_eq_var}), which is numerically
small under typical experimental conditions, one has that
\begin{eqnarray}
k&=&\omega\frac{m^{1/2}}{\omega_{r,0}^{1/2}\hbar^{1/2}}
(2a_{s}\rho)^{-1/2}(1+2a_{s}\rho)^{3/4}
\nonumber
\\[1.0ex]
&&\times (4+6a_{s}\rho)^{-1/2},
\label{eq_disp_anal}
\end{eqnarray}
where $\rho=N/2L$ is the density of the condensate.

It is important to note that the above formula
for $k$ [Eq.~(\ref{eq_disp_anal})] has been obtained
by assuming a homogeneous condensate with constant
density $\rho=\rho_0$. This approximation yields a spacing
between adjacent maxima of
\begin{equation}
S_0 = \frac{2 \pi}{k},
\label{S0}
\end{equation}
where $k$ is computed using Eq.~(\ref{eq_disp_anal}) with
$\rho=\rho_0$.
However, the density of the condensate in the considered system is
not homogeneous. To account for this inhomogeneity, as a
first-order approximation, one  uses the fact that, typically for
the cases under consideration, the density of the condensate
varies on a space scale much larger than that of the observed
patterns and thus we can extend Eq.~(\ref{eq_disp_anal}) with the
density being space dependent: $\rho=\rho(x)$.
In fact, the density of the condensate can be approximated
in this slowly-varying limit by
the so-called Thomas-Fermi (TF) approximation:
\begin{equation}
\rho(x)=3\frac{L^{2}-x^{2}}{4L^{3}},
\label{rhoTF}
\end{equation}
when $-L<x<L$ and $\rho(x)=0$ otherwise. Therefore, it is
possible to approximate the average spacing by taking
a spatially averaged $\bar{k}$ for $k$ in Eq.~(\ref{eq_disp_anal}) using
$\rho(x)$ given by the TF approximation in Eq.~(\ref{rhoTF}):
\begin{equation}
{\cal S}_{1}=\frac{2\pi}{\bar{k}},
\label{S1}
\end{equation}
where
\begin{equation}
\bar{k}=\frac{1}{2L}\int_{-L}^{L}k(x)dx.
\end{equation}
Alternatively, one could average the spacing directly by using
the expression
\begin{equation}
{\cal S}_{2}=\frac{1}{2L}\int_{-L}^{L}\frac{2\pi}{k(x)}dx,
\label{S2}
\end{equation}
where, again, $k(x)$ is given by Eq.~(\ref{eq_disp_anal})
with the Thomas-Fermi density $\rho(x)$ of Eq.~(\ref{rhoTF}).

\begin{figure}[tbhp!]
     \centering
           \includegraphics[width=8cm]{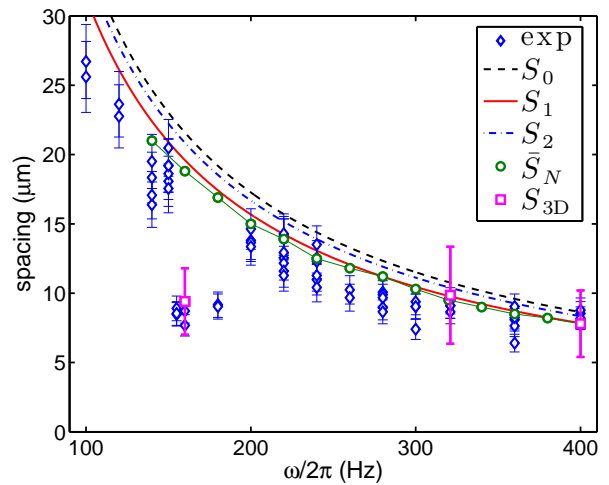}
      \caption{(Color online)
Average spacing of adjacent maxima of the longitudinal patterns as a
function of the transverse driving frequency. Blue diamond dots
depict the experimental data of Ref.~\cite{engels}
(the experimental error bar depicted here only takes into account the
error bar in the pixel size of the recorded experimental image).
Dashed black
line, solid red line and blue dashed-dotted line correspond,
respectively, to spacings computed using $S_0$, $S_1$ and $S_2$.
Green empty circles correspond to spacings extracted from the 1D
NPSE numerics by the averaged spacing method. Pink squares
correspond to the spacings extracted from the full 3D numerics
using the FFT method and its associated error bars (see
text for details). Notice that for $\omega/2\pi$ close to 160.5 Hz
the theoretical prediction is far from the experimental data.
The radial breathing mode excited at these frequencies
cannot be captured by the NPSE but is well captured by the 3D numerics.}
      \label{fig_e1}
\end{figure}

The above expressions for the spacing $S$ provide the analytical
prediction of the present study that can be readily compared
quantitatively with the experimental results of
Ref.~\cite{engels}. This comparison can be seen in
Fig.~\ref{fig_e1}, where the theoretical predictions for $S_0$,
$S_1$ and $S_2$ defined above are adapted to the $^{87}$Rb
experiments of Ref.~\cite{engels}, with
$\omega_{r,0}/(2\pi)=160.5$~Hz, the condensate length $2 L =
180~\mu$m and $N=5 \times 10^5$ atoms \cite{cond_length}. We
observe a very good qualitative and a good
quantitative agreement between the theoretical
predictions and the experimental result, solidifying our
expectation that Eq.~(\ref{eq_disp_anal}) captures accurately and
in a fully analytical way the observed phenomenology of the
experiment of Ref.~\cite{engels}. Is it worth mentioning that
$S_1$ and $S_2$ are closer to the experimental data since they
account for the inhomogeneity of the cloud. Notice as well that
our analytical prediction shows deviations from the experimental
data at low frequencies, where the spatial periods of the Faraday
waves are comparable with the length of the condensate. At larger
frequencies we have good agreement between the theoretical curve
and the experimental data, as the periods of the Faraday waves are
substantially smaller than the spatial extent of the cloud. The main
sources of slight disparity between the theoretical predictions
and the experimental results can be traced in the transverse Gaussian
(as opposed to Thomas-Fermi) profile and the fact that the analysis
cannot directly incorporate the weak longitudinal trapping potential
(see the discussion above). These will be further clarified below,
through the comparison with the direct numerical simulation
results of both the NPSE as well as the full 3D GP equation.

\begin{figure}[tb]
\centering
\includegraphics[width=7cm,height=5cm]{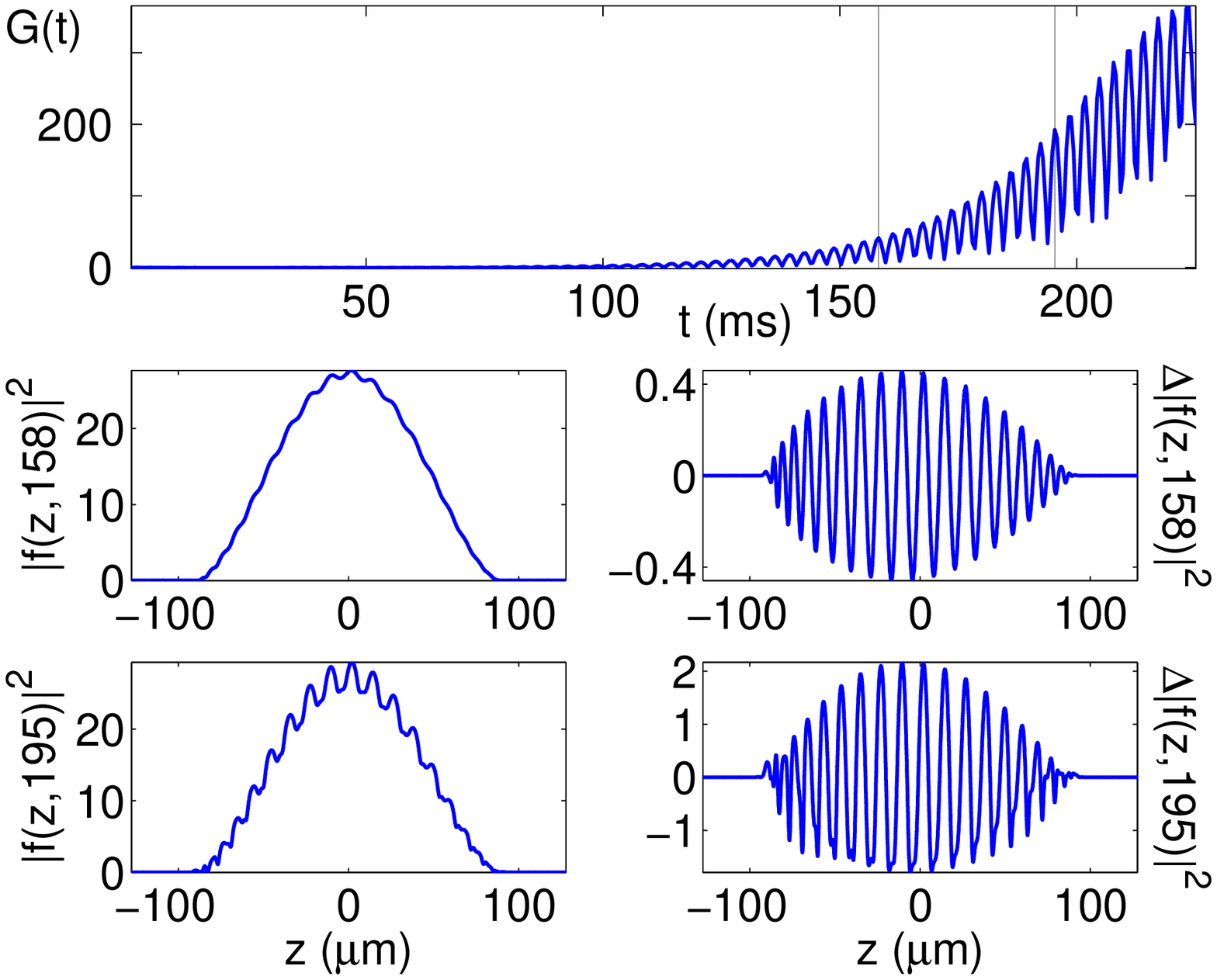}
\\[1.0ex]
\includegraphics[width=8cm]{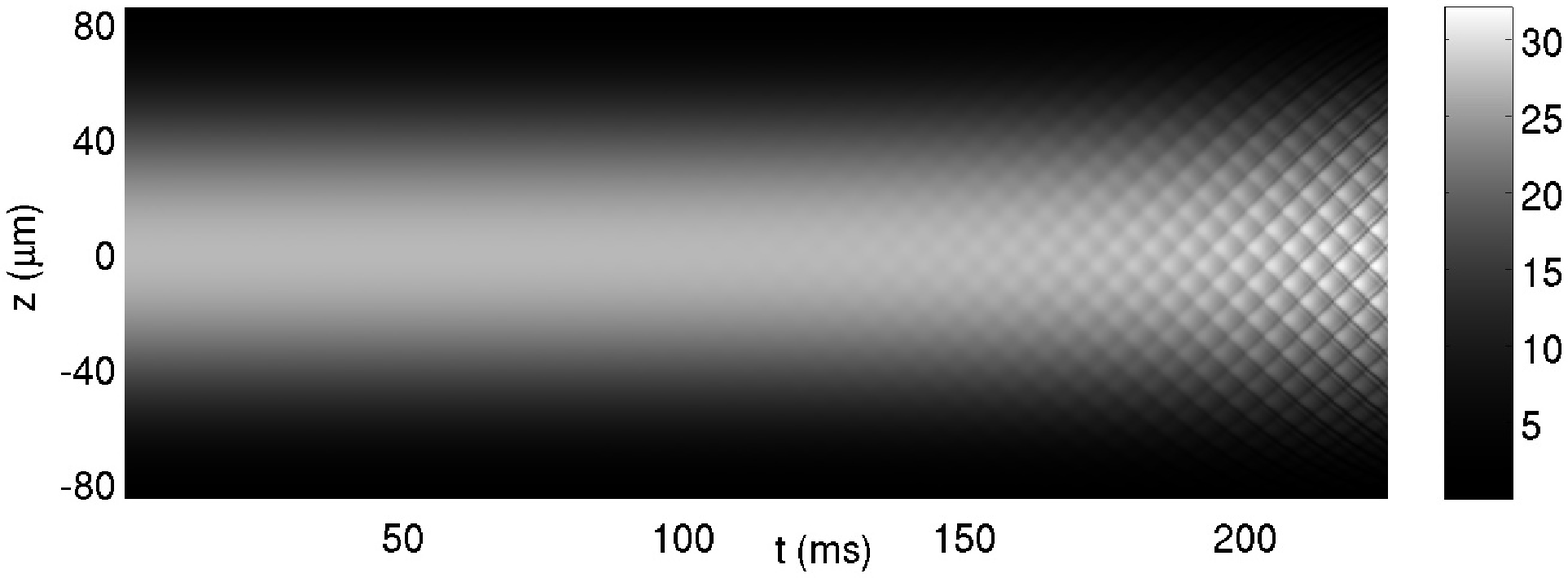}
\caption{Faraday pattern formation for the 1D NPSE.
The top panel depicts the growth rate $G$ ($L^2$
norm of the deviation from the initial density) of the pattern.
The middle two rows display the Faraday patterns
at the times indicated (see vertical lines in top panel)
where the left subpanels show the density profile while the
right subpanels show the deviation from the initial density
profile. The bottom panel depicts the space-time evolution
of the density. This case corresponds to the
experiment in Ref.~\cite{engels}, namely, a cloud
of $N=5 \times 10^5$ $^{87}$Rb atoms, trapped by
$\{\omega_{r,0},\omega_z\}/(2\pi)=\{160.5,7\}$~Hz
with a 20\% modulation of the radial confinement at
a frequency $\omega/(2\pi)=321$~Hz.
}
\label{fig1d1}
\end{figure}

\section{Numerical Results}
\subsection{One-Dimensional Numerics on the NPSE}

Having completed the linear stability analysis, let us now
turn to full numerical simulation to investigate the instability
onset and the emergence of the relevant Faraday patterns.
We have simulated the NPSE (\ref{salasnich_eq}) for the
experimental conditions described in Ref.~\cite{engels}.
Specifically, in Fig.~\ref{fig1d1}, we show the formation of
the Faraday pattern for a condensed cloud of
 $N=5 \times 10^5$ $^{87}$Rb atoms contained in a magnetic trap with
frequencies $\{\omega_{r,0}/(2\pi),\omega_z/(2\pi)\}=\{160.5,7\}$~Hz
where the radial trap frequency has a modulation of 20\% ($\epsilon=0.2$,
which is within the typical range of experimentally used modulations)
and a frequency $\omega/(2\pi)=321$~Hz corresponding to the
resonant oscillation frequency of the radial breathing mode
(i.e., $\omega \approx 2\omega_{r,0}$) \cite{string1,perez2}.
As it can be observed from the figure, the Faraday pattern grows
exponentially until it is clearly visible in the density space-time
evolution (bottom panel) after about 125~ms. It is reassuring that
the NPSE is successful in capturing the Faraday pattern with
the same wavelength of 10--11~$\mu$m as the
experiment of Ref.~\cite{engels}. On the other hand, we have found that
the NPSE cannot capture the right time for the development of
the instability. The NPSE results take about 125~ms (i.e.,
about 40 periods of the modulation drive) for the
Faraday pattern to be visible while, in the experiments of
Ref.~\cite{engels}, some of the patterns are clearly visible after
some 10 periods of the drive.

\begin{figure}[tb]
\centering
\includegraphics[width=7cm,height=5cm]{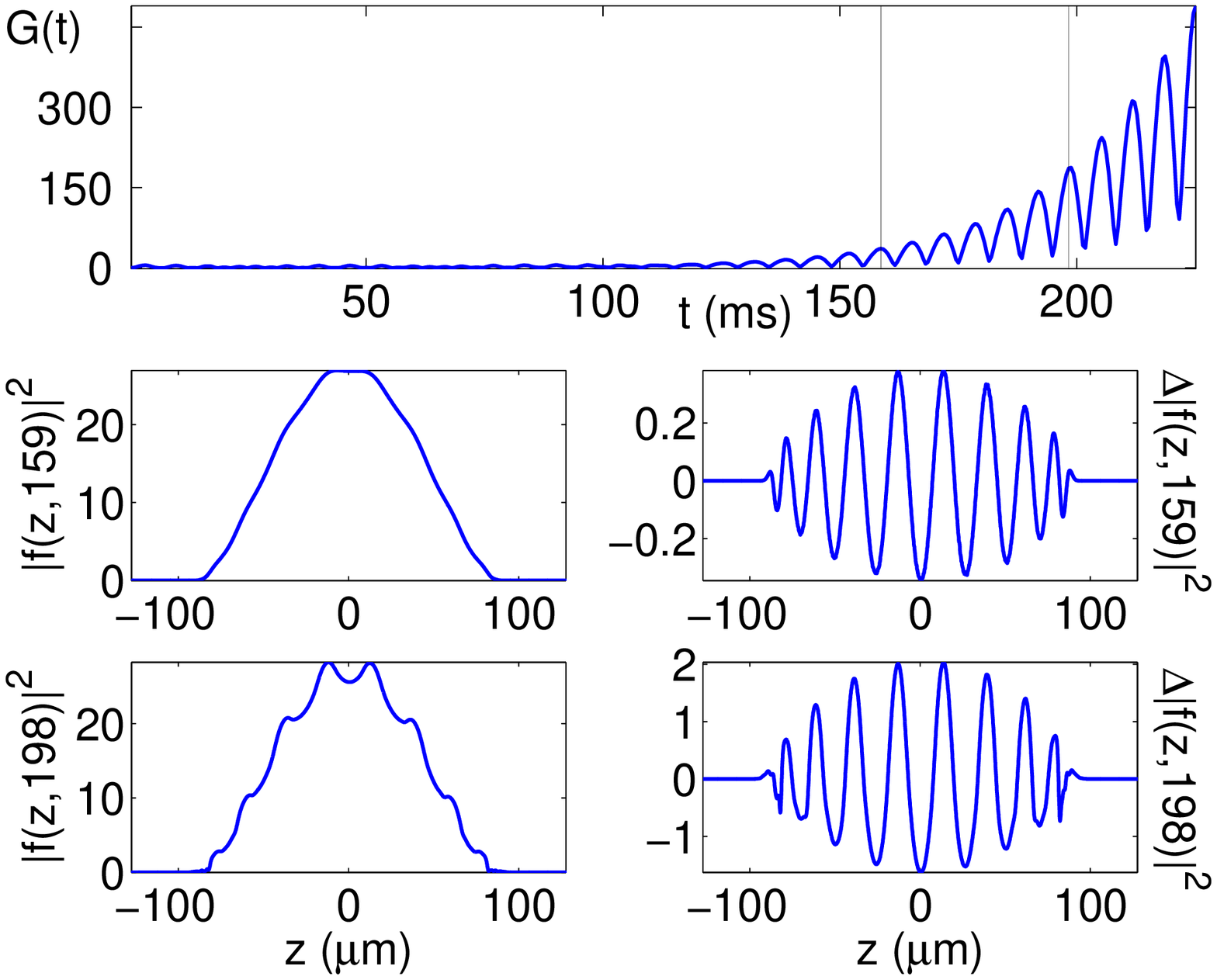}
\\[1.0ex]
\includegraphics[width=8cm]{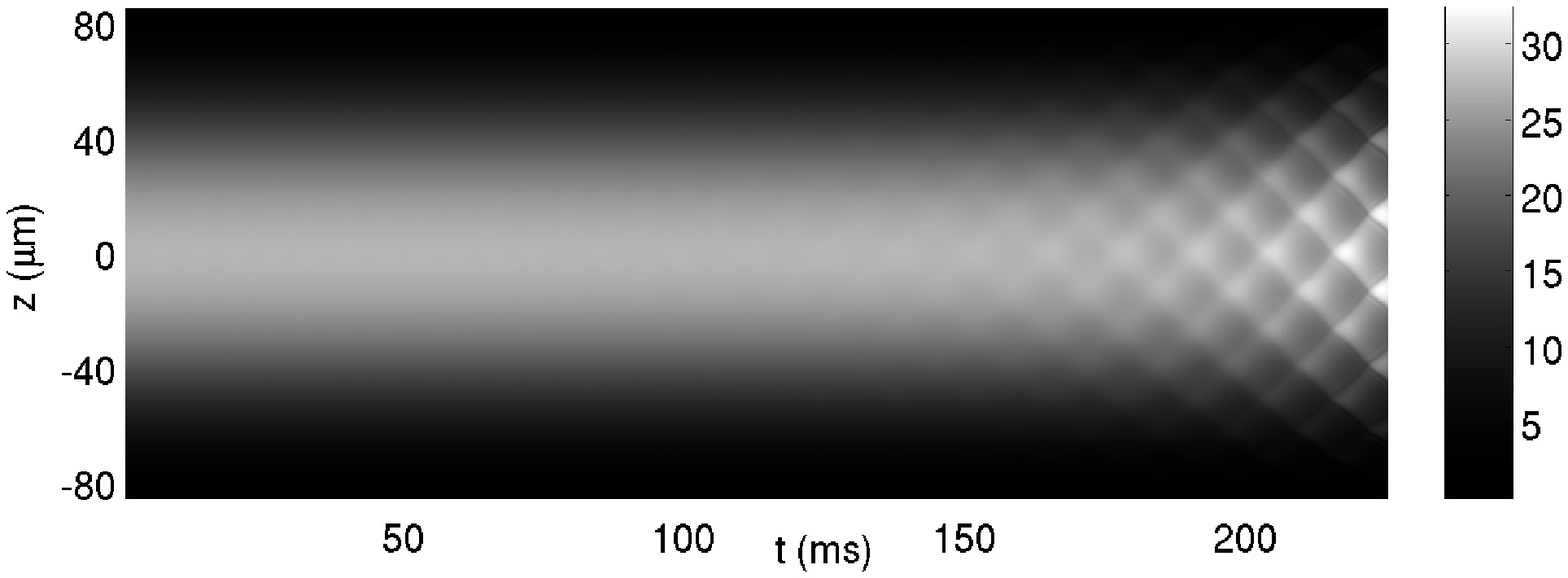}
\caption{Same as in Fig.~\ref{fig1d1} for a
(stronger) 40\% modulation of the radial confinement at
a (slower) driving frequency $\omega/(2\pi)=150$~Hz.
}
\label{fig1d2}
\end{figure}

A possible explanation for the discrepancy of the Faraday
pattern growth between NPSE and the experiments may lie
in the size of the initial perturbation,
that will eventually seed the Faraday pattern.
We have used various
amplitudes for the initial perturbation after we
obtained the steady state solution to Eq.~(\ref{salasnich_eq})
by imaginary time relaxation. In the results presented
in this work we used an initial perturbation with an amplitude
randomly chosen in an interval 0.001 times the local density.
We also tried larger perturbations, up to ten times larger,
and the effect is to accelerate the appearance of the
patterns (results not shown here), however we were unable
to see distinguishable patterns earlier than 30--35
driving periods for the above setting.
Another effect that needs to be taken
into account is the amplitude of the modulation drive.
While in the experiments of Ref.~\cite{engels} Faraday
patterns, for the resonant frequency $\omega/(2\pi)=321$~Hz,
quickly formed for even small drive amplitudes (less than 4\%, i.e.,
$\epsilon<0.04$), our numerical results using the NPSE
always needed a much larger drive amplitude ($\epsilon=0.2$
in the results of Fig.~\ref{fig1d1}). Therefore, it is
clear that the NPSE, although clearly able to capture the
nature (wavelength) of the Faraday pattern, it is unable to
predict the growth rate of the instability. The reason for this
shortcoming stems from the fact that the transverse component of
the wavefunction in Eq.~(\ref{gaussian_ansatz_var})
is considered to be at its {\em ground state} at {\em all times}.
This is quite a strong assumption considering that the
cloud presents impact oscillator-type dynamics for its
radial width \cite{engels} (cf.~top panel of Fig.~\ref{fig3d1}
below).
This will be verified when we relax the
radial wavefunction profile in the 3D simulations
shown below.

\begin{figure}[t]
\begin{center}
\includegraphics[width=8.5cm]{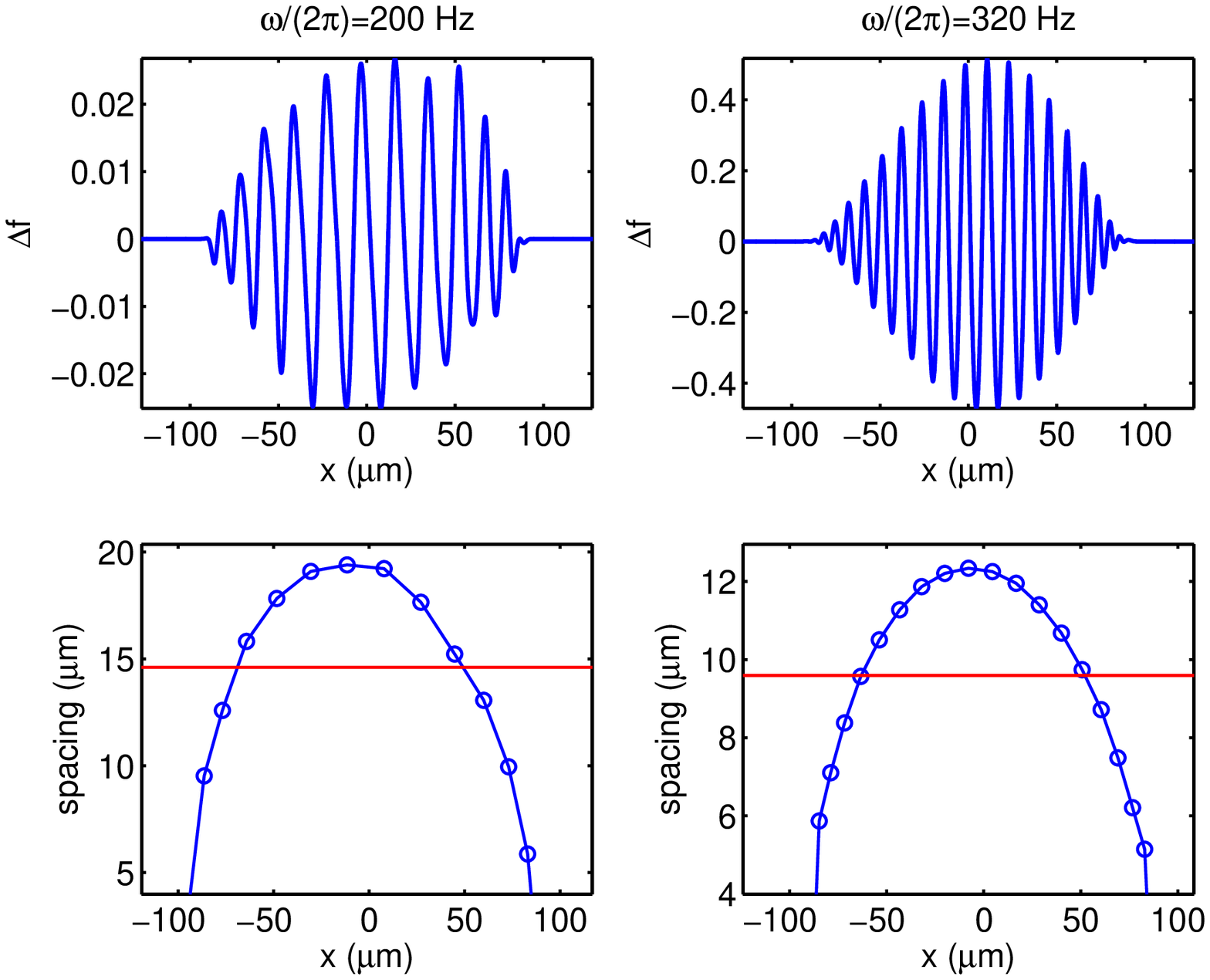}
\end{center}
\vspace{-0.5cm}
\begin{center}
~~\includegraphics[width=8.05cm]{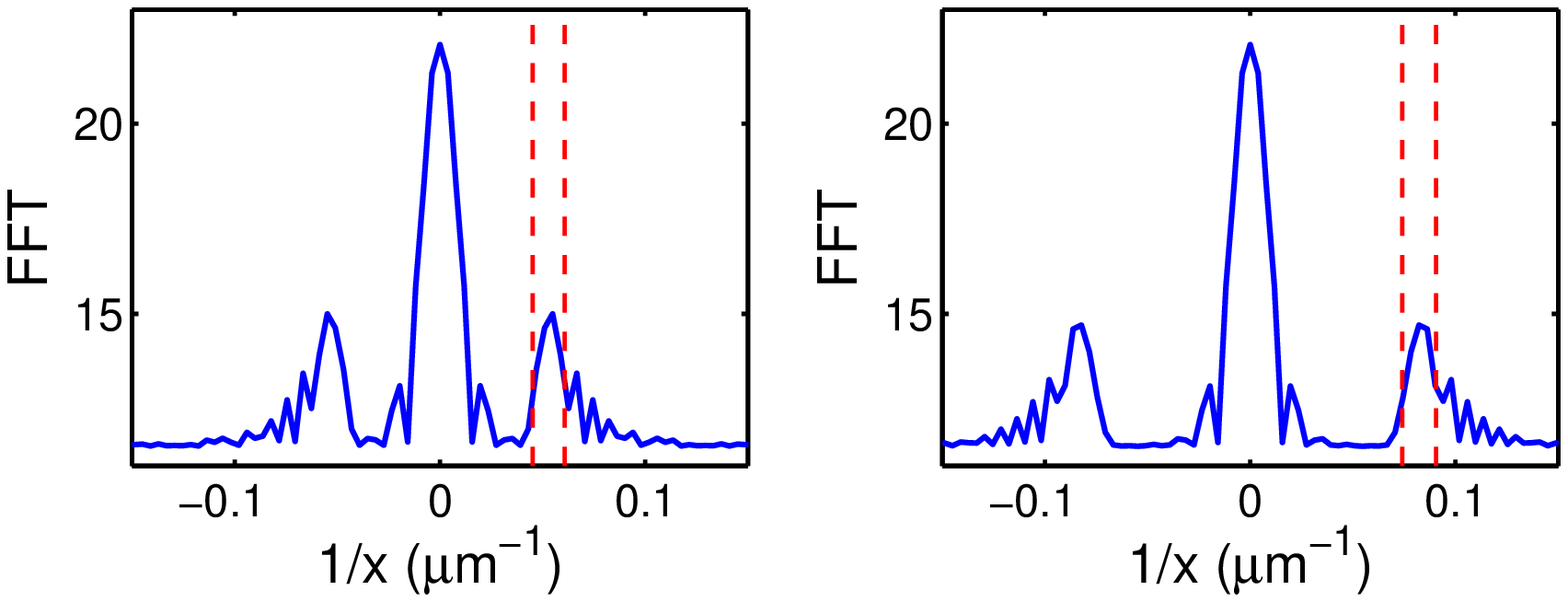}
\end{center}
\caption{(Color online) Method of averaging spacings in the
inhomogeneous cloud for the NPSE model. The left and right columns
of plots correspond, respectively, to $\omega/(2\pi)=200$~Hz and
$\omega/(2\pi)=320$~Hz. The top row of panels shows the density
deviation $\Delta f$ from the initial density profile. The middle
row of panels depicts with dots the measured spacing between
maxima of $\Delta f$ inside the cloud. The averaged spacing is
shown with the red horizontal line. The bottom row of panels shows
the FFT of the density $|f|^2$. The vertical lines depict the
estimated window of the frequencies contained in the Faraday
pattern due to the non-homogeneity in the spacings (see middle
row). } \label{spacings_NPSE}
\end{figure}

The results presented in Fig.~\ref{fig1d1} correspond to
the most pattern-forming sensitive case since the
drive of the radial frequency is tuned to resonate
with the natural breathing frequency of the radial
mode. For other driving frequencies, the growth of the
Faraday pattern is less pronounced. This is demonstrated
in Fig.~\ref{fig1d2}, where we use the same parameter
values as in Fig.~\ref{fig1d1} but changed the driving
frequency to $\omega/(2\pi)=150$~Hz and we
doubled its amplitude ($\epsilon=0.4$). For this
out-of-resonance frequency, the Faraday pattern takes longer
to form and even a drive with twice the amplitude
takes longer to seed the pattern (the pattern is not
visible until approximately 140ms). Nonetheless, it is
interesting that this out-of-resonance case only takes
about 20 periods to manifest itself.

To summarize the results of the 1D NPSE simulations and to compare
them with our analytical prediction for the spacing of the ensuing
Faraday patterns, we proceed to measure the averaged spacing in
the simulations. The method relies on computing the spacing
between maxima on the density deviation $\Delta f$ from the
initial profile (see top row in Fig.~\ref{spacings_NPSE}). A
couple of examples of the dependence of the spacing inside the
cloud are depicted in Fig.~\ref{spacings_NPSE} together with their
average, from now on denoted as $\bar{S}_N$. It is clear from
these examples that at the center of the cloud, where the density
is larger, the spacing is larger than at the periphery of the
cloud where the density is lower. The averaged spacing
$\bar{S}_N$ was computed as a function of the driving frequency
and it is depicted by the green empty circles in
Fig.~\ref{fig_e1}. As it is clear from Fig.~\ref{fig_e1}, the
spacing $S_1$ [computed using the analytical expression for the
spacing given in Eq.~(\ref{eq_disp_anal}) with a spatially
averaged density on the TF approximation] and the averaged spacing
$\bar{S}_N$ from the NPSE dynamics are in good agreement (and the
relevant approximation of using Eq.~(\ref{eq_disp_anal}) together
with Eq.~(\ref{rhoTF}) to represent the effects of the
longitudinal potential is a fairly accurate one).
It is important to mention that the non-homogeneity of the
spacings induces an inherent uncertainty in the quantification of
the associated spacing for a particular Faraday pattern. The
associated window of spacings contained in the Faraday pattern can
also be seen from the fast Fourier transform (FFT) spectrum of the
density. In the bottom panels of Fig.~\ref{spacings_NPSE} we
depict the FFT spectrum of the density for a couple of cases with
their respective frequency windows (see below for further
elaboration on this effect).

\subsection{Three-Dimensional Numerics}

\begin{figure}[tb]
\centering
\includegraphics[width=6.8cm,height=1.75cm]{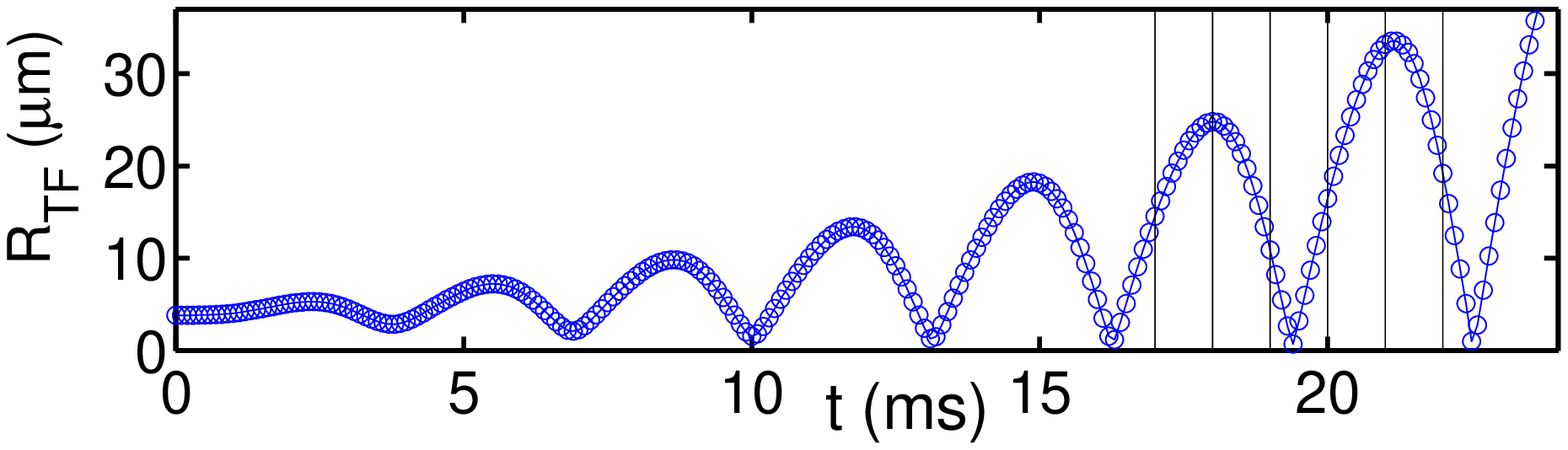}
\includegraphics[width=7cm,height=5cm]{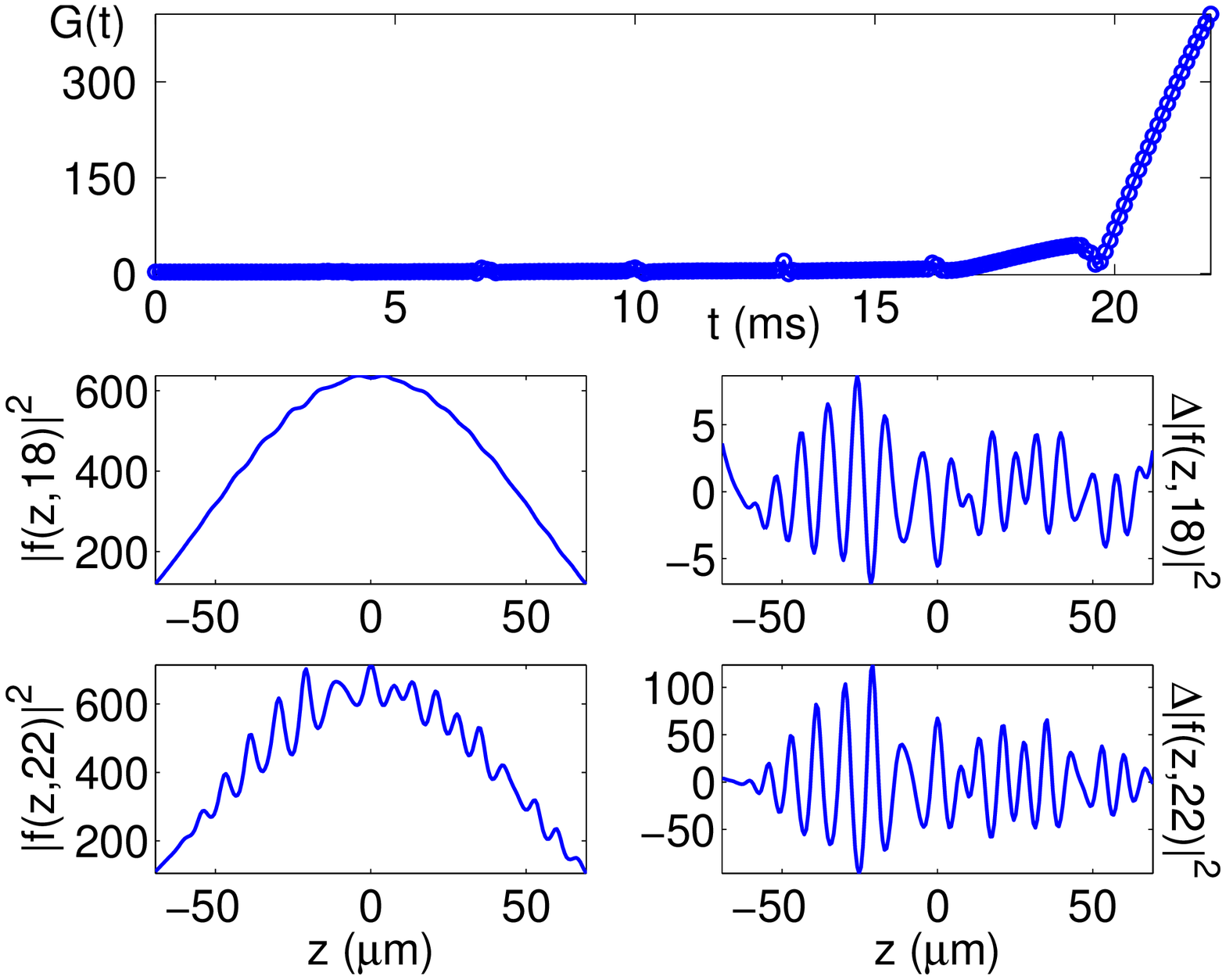}
\caption{Faraday pattern from the $(r,z)$ 3D simulations.
This case corresponds to the
experiment in Ref.~\cite{engels}, namely, a cloud
of $N=5 \times 10^5$ $^{87}$Rb atoms, trapped by
$\{\omega_{r,0},\omega_z\}/(2\pi)=\{160.5,7\}$~Hz
with a 20\% modulation of the radial confinement at
a driving frequency $\omega/(2\pi)=321$~Hz.
The top panel depicts the transverse radius of the
cloud displaying impact-oscillator behavior
(thin vertical lines depict the times of the
snapshots shown in Fig.~\ref{fig3d2}).
The second panel depicts the growth of the
Faraday pattern while the bottom two rows
depict the $r$-integrated density profiles
(left subpanels) and their deviation from the
initial profile (right subpanels).
}
\label{fig3d1}
\end{figure}

\begin{figure}[tb]
\centering
\includegraphics[width=8cm,height=3.3cm]{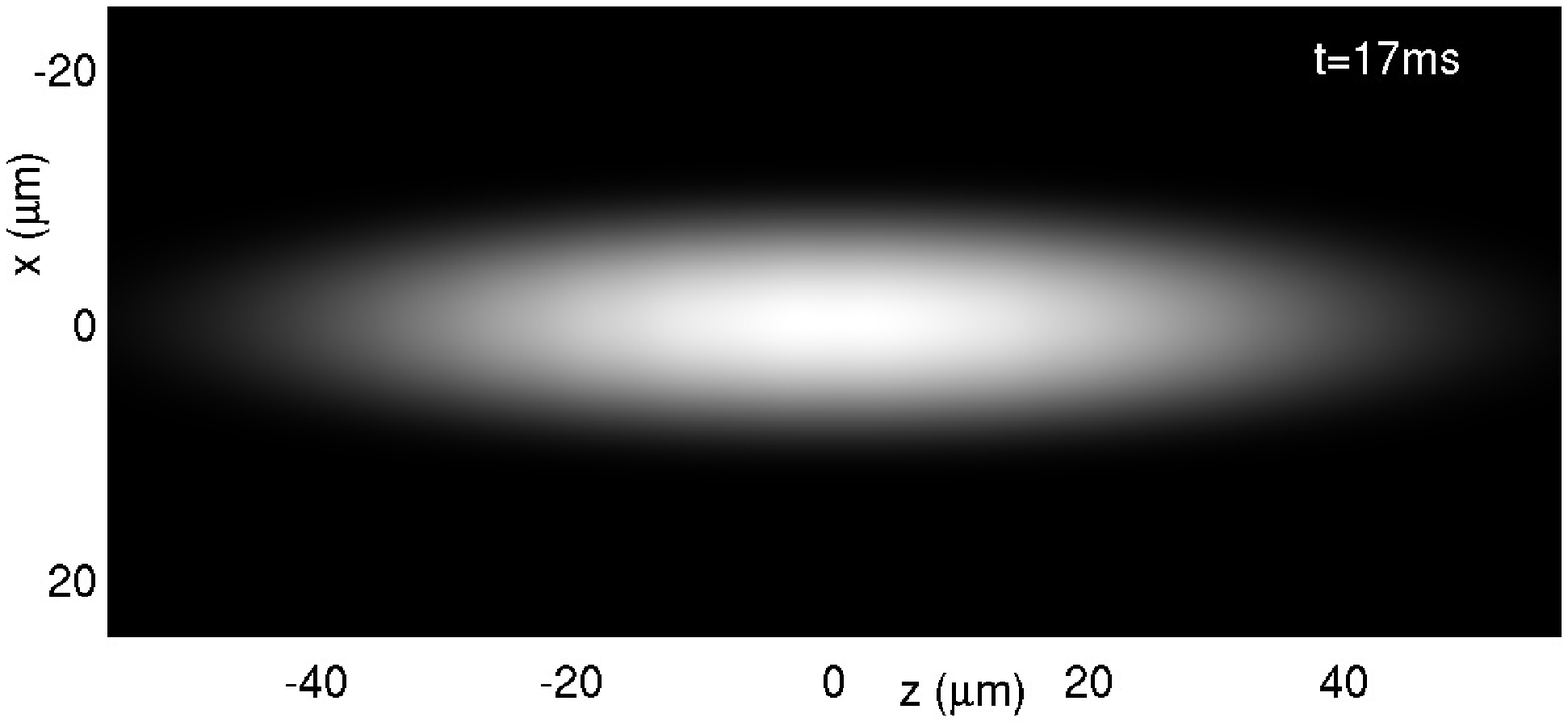}
\\[-2.5ex]
\includegraphics[width=8cm,height=3.3cm]{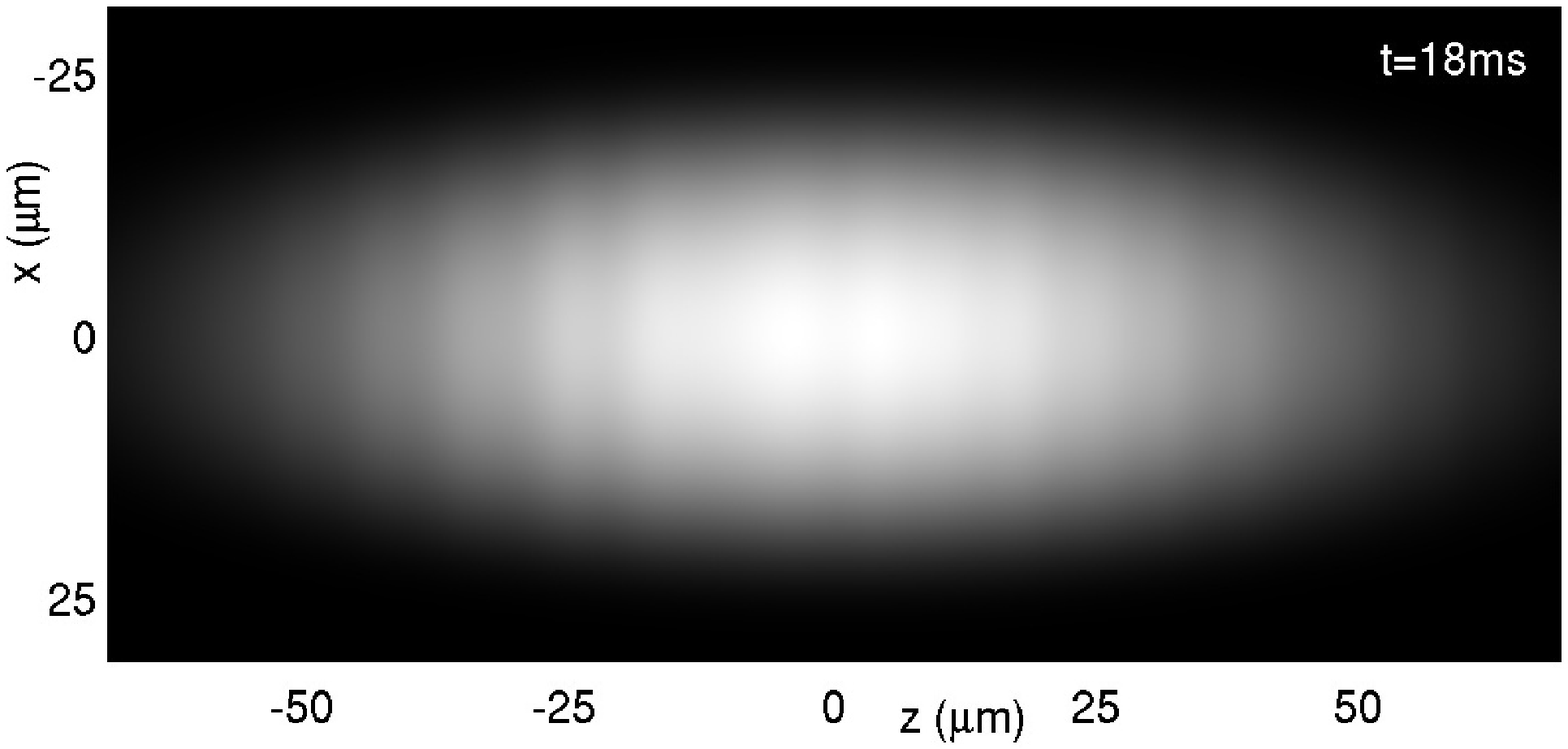}
\\[-2.5ex]
\includegraphics[width=8cm,height=3.3cm]{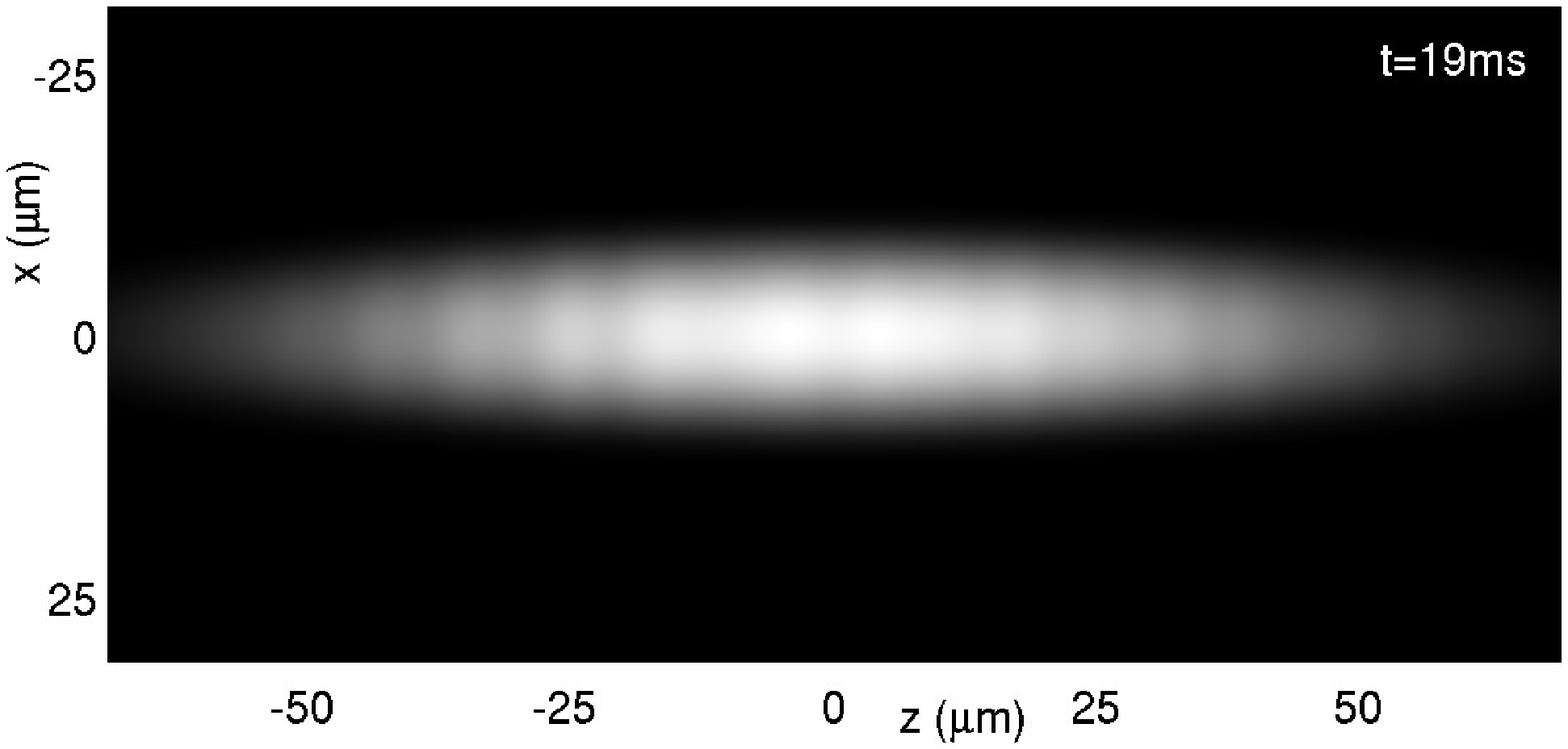}
\\[-2.5ex]
\includegraphics[width=8cm,height=3.3cm]{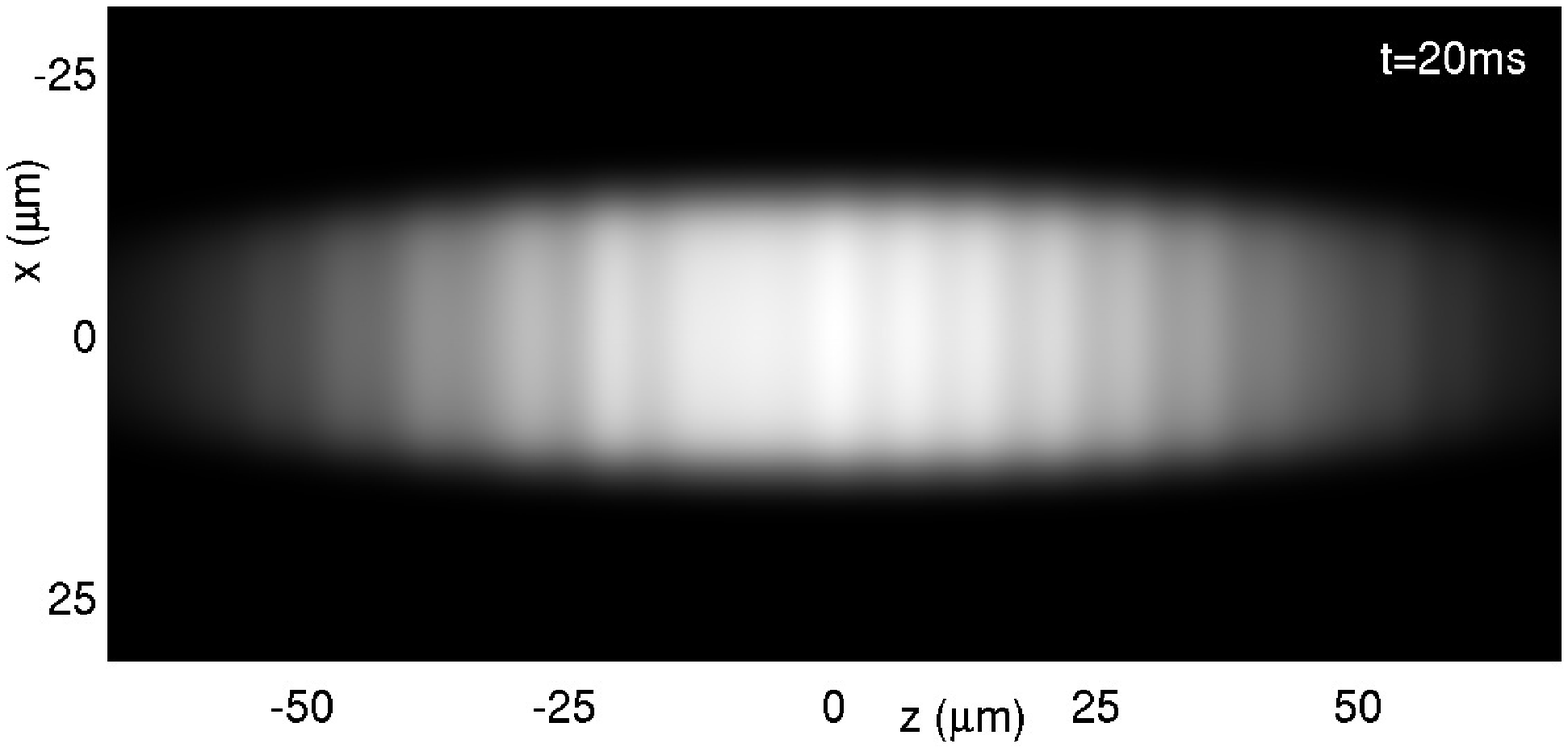}
\\[-2.3ex]
\includegraphics[width=8cm,height=3.3cm]{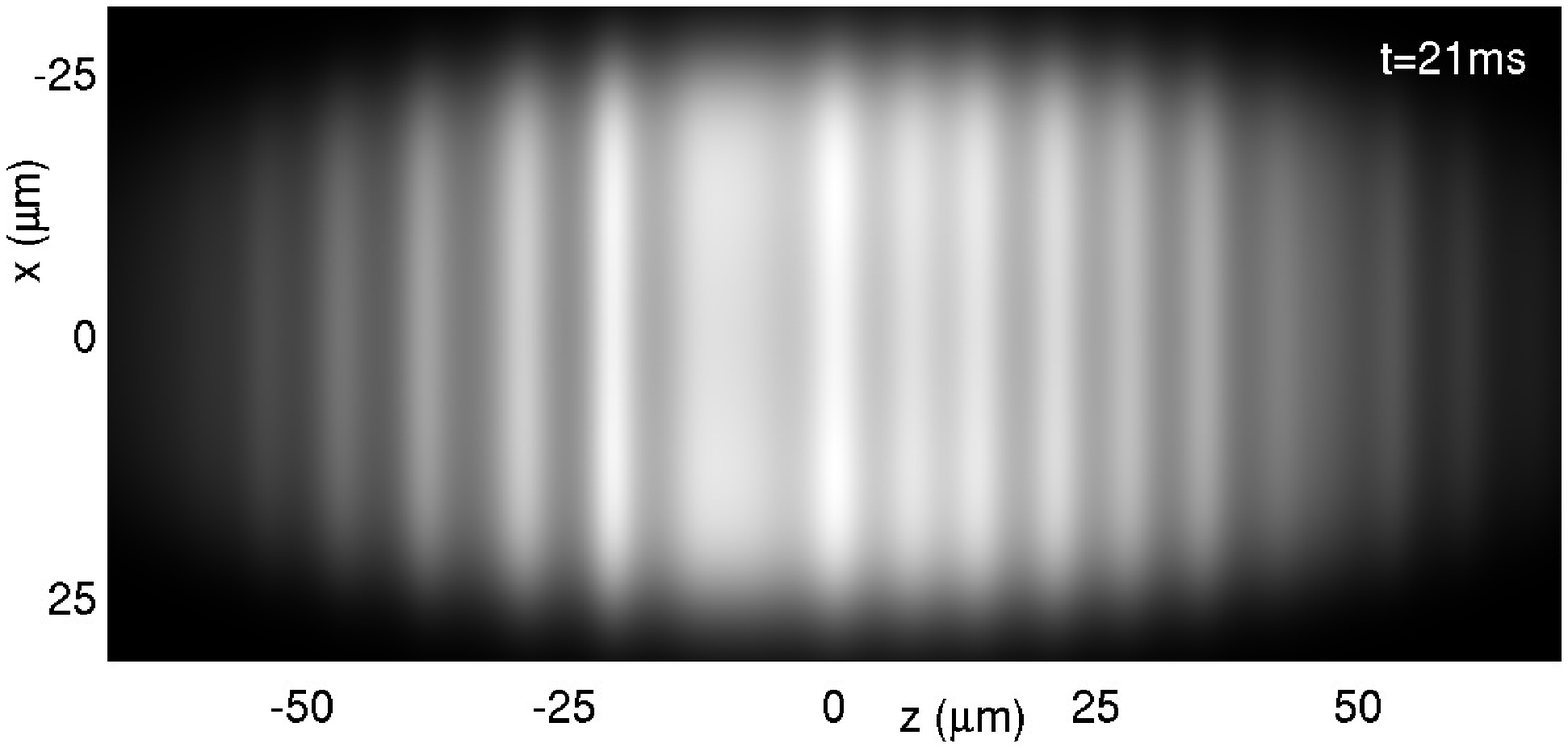}
\\[-2.3ex]
\includegraphics[width=8cm,height=3.3cm]{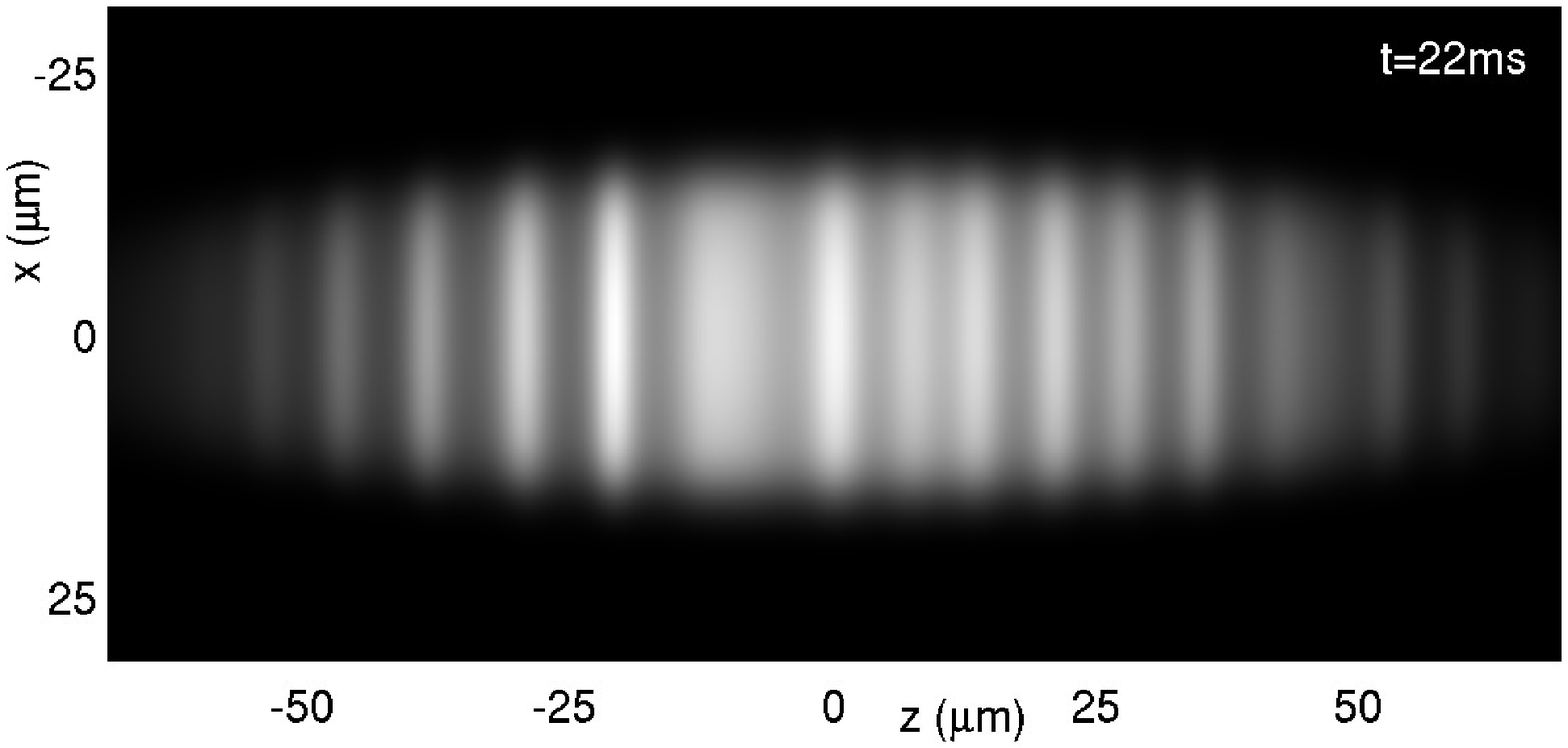}
\caption{Development of the Faraday pattern in our
3D numerical simulations.
Shown are the snapshots of the $y$-integrated profile density
(i.e., the observable in the experiments)
for the same data as in Fig.~\ref{fig3d1} at
the times indicated.
}
\label{fig3d2}
\end{figure}

In order to more accurately model the Faraday pattern
formation, we used direct numerical simulations of the
Gross-Pitaevskii equation (\ref{GPE}). The numerics consist
of integrating Eq.~(\ref{GPE}) in cylindrical coordinates.
The choice of cylindrical symmetry instead of full 3D numerics
is justified by the fact that the radial direction
does not develop azimuthal instabilities as it can
be observed in the experiments of Ref.~\cite{engels}
(and it was also verified by additional tests runs
of the full 3D equation on a coarser grid).
The main challenge in numerically integrating
Eq.~(\ref{GPE}) stems from the impact oscillator-type
dynamics of the radial profile that is driven at
resonance. These oscillations produce two
numerically challenging effects:
(a) they bring most of the atoms close to $r=0$ requiring
   an extremely fine grid, and
(b) as the cloud accelerates during the impact oscillations,
   the wavefunction oscillates, in space, very rapidly (although the
   density does not) again requiring a very fine grid.
Therefore, although (a) could be circumvented by a
grid refinement around $r=0$, challenge (b) requires
a fine $r$-grid where the cloud is traveling fastest
and this happens on a large portion of the domain. Thus a fine grid
needs to be implemented throughout the (radial) $r$-direction,
and as a (numerical scheme stability)
consequence a very small time step is also required.
For the (longitudinal) $z$-direction, it suffices to have
enough points to accurately capture the Faraday pattern
whose wavelength is quite manageable. In the 3D simulations
shown in Figs.~\ref{fig3d1} and \ref{fig3d2} we were able to
accurately integrate Eq.~(\ref{GPE}) for about 7 cycles
of the drive with a grid of 2001$\times$401 points
in the $(r,z)$-plane with a finite difference scheme in
space with 4--5$^{\rm th}$ Runge Kutta in time with
a time step of $0.00025$ (in adimensionalized units).

The initial condition used in the simulations (i.e., the ground
state of the condensate) was obtained, as for the 1D case, by
imaginary time relaxation.

Figures~\ref{fig3d1} and \ref{fig3d2} depict the
Faraday pattern arising
from the $(r,z)$ Gross-Pitaevskii simulation for a cloud of
$N=5 \times 10^5$
$^{87}$Rb atoms driven at resonance ($\omega/(2\pi)=321$~Hz)
by a modulation amplitude of 20\%, corresponding to
the experiments of Ref.~\cite{engels}.
Depicted in  Fig.~\ref{fig3d2} is the $y$-integrated (top)
view which is what is measured in the experiments. Clearly
observable are the well separated fringes of the forming
Faraday pattern with an approximate spacing of about 8.5~$\mu$m.
Furthermore, in contrast with the NPSE simulations, the Faraday
instability develops more rapidly (see second panel in
Fig.~\ref{fig3d1}) and it is clearly observable after only 6--7
periods of the drive. As mentioned above, the instability sets in
much more rapidly in the full 3D system than in the 1D NPSE
reduction, as expected based on the previous discussion. It is
worth mentioning that in the 3D simulations we did not introduce a
perturbation to the initial condition to seed the Faraday patterns
since the inherent numerical noise was capable of starting the
pattern.

\begin{figure}[tb]
\centering
\includegraphics[width=8.5cm]{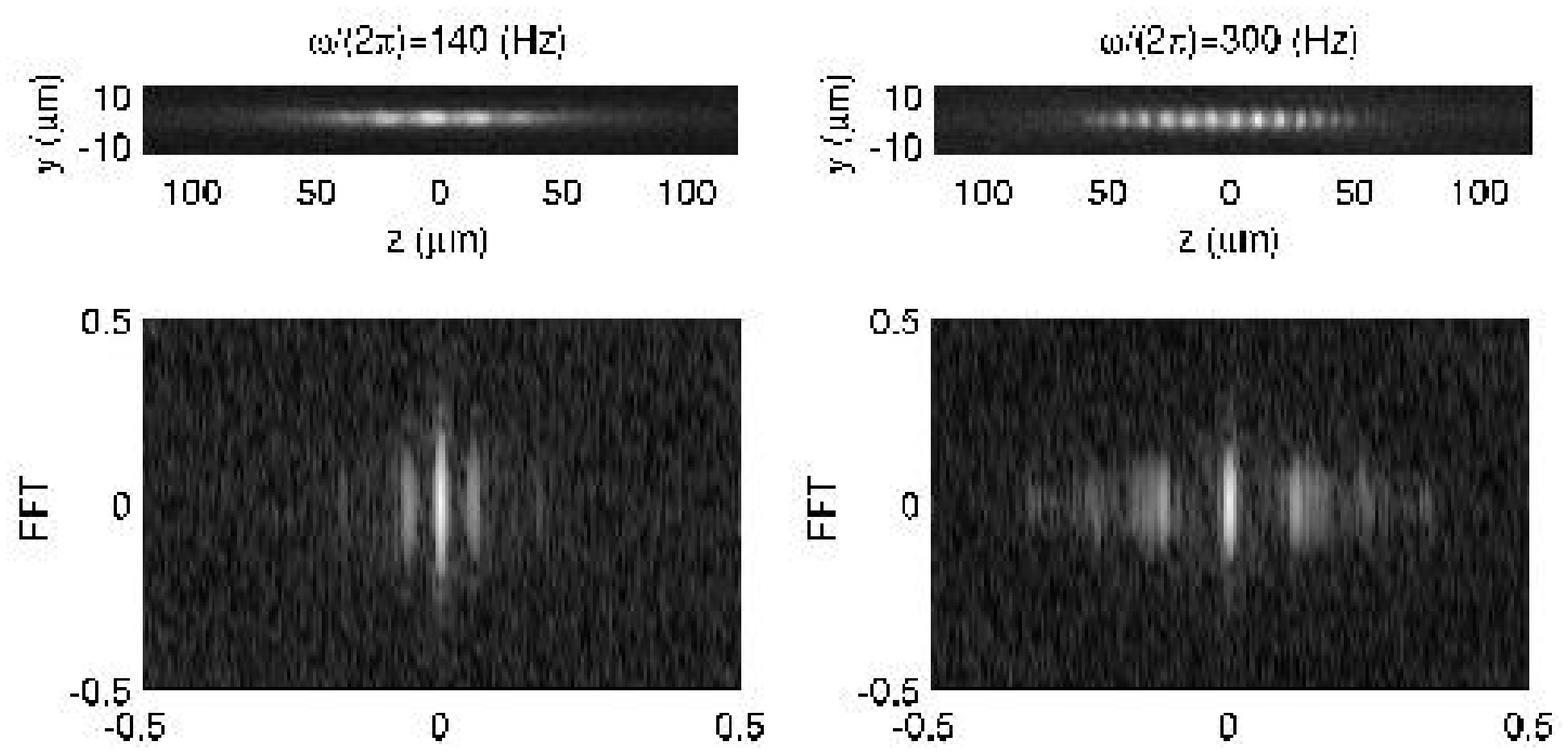}
\\[2.0ex]
\hskip-0.2cm
\includegraphics[width=8.7cm]{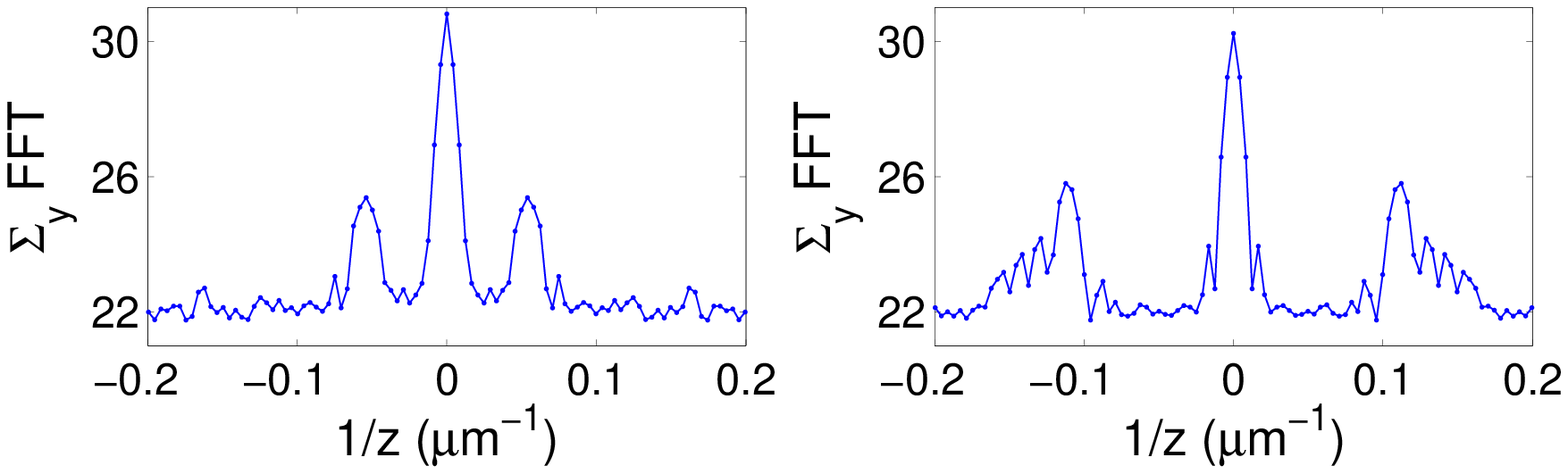}
\caption{(Color online)
FFT analysis for the data from the experiments of Ref.~\cite{engels}.
Left and right column correspond, respectively, to a driving
frequency of $\omega/(2\pi)=140$~Hz and $\omega/(2\pi)=300$~Hz.
The top row depicts the images from the experiment.
The second row depicts the corresponding FFTs and the third
row the FFTs integrated over the $y$ direction.
}
\label{FFTexp}
\end{figure}

\begin{figure}[tb]
\centering
\includegraphics[width=6.5cm]{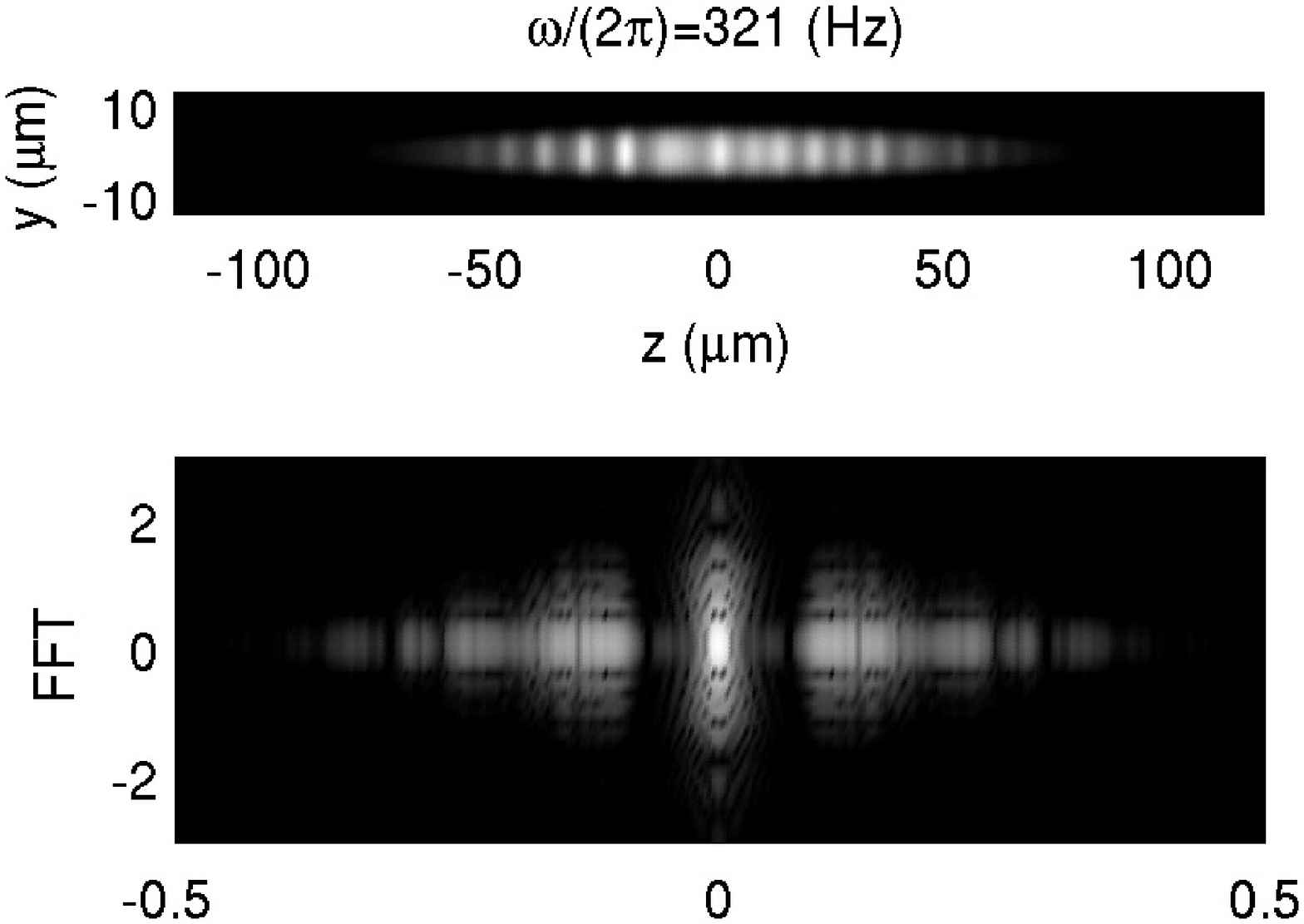}
\\[1.5ex]
~\includegraphics[width=6.6cm]{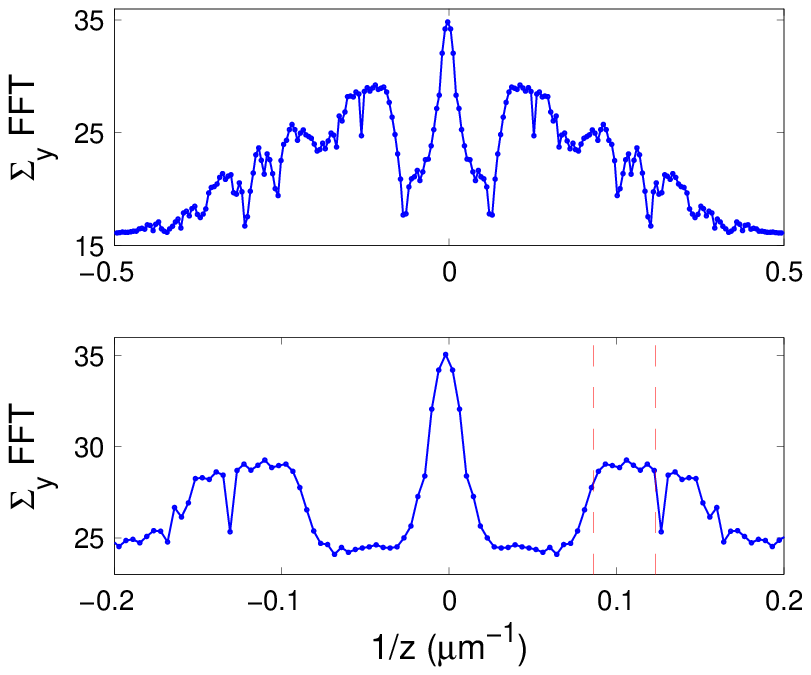}
\caption{(Color online)
FFT analysis from the 3D numerics for $\omega/(2\pi)=321$~Hz.
The top two panels depict, respectively, the Faraday pattern
and its FFT.
The third panel depicts the FFT integrated
over the $y$ direction. The bottom panel depicts the
same as the third panel but the FFT is computed
after adding a 5\% noise to the density to simulate the
experimental noise in the picture. The vertical lines
in the last panel represent our estimated window of
possible frequencies for the Faraday pattern (see error
bars in the pink square points in Fig.~\ref{fig_e1}).
}
\label{FFTnum}
\end{figure}

To validate our full 3D numerics we have computed the
average spacing (using the FFT method, see below)
for three different cases of the driving
frequency. The results are depicted by the pink squares
in Fig.~\ref{fig_e1}. As it is evidenced from the figure,
the 3D numerical simulations accurately reproduce the spacing
from the experiments in Ref.~\cite{engels} including
the case when the external driving frequency is {\em half}
of the resonant frequency (see left-most pink square point
in Fig.~\ref{fig_e1}).

In Ref.~\cite{engels}, the sudden drop in the spacing around
$\omega/(2\pi)=160$~Hz, i.e., half of the resonant frequency, was
attributed to excitation of the radial breathing mode. However,
our numerics suggest that this is not the case and that this is
due to the fact that we are driving a sub-harmonic that excites
the resonance.

It is important to note that the data for the experiment in
Ref.~\cite{engels} has a significant variability in the
spacing values. This natural experimental variability has
many potential sources. Here, we would like to focus on
the error generated by the width of the frequency peak
used to measure the spacing through the fast Fourier
transform (FFT). In the experiments of Ref.~\cite{engels}
(see blue diamond dots in Fig.~\ref{fig_e1}), the spacing
of the Faraday pattern
was measured by computing the FFT of the integrated density
image and extracting the spatial frequency of the dominant peak.
The error bars depicted in Fig.~\ref{fig_e1} for the
experimental data {\em only} take into account the error bar in the
pixel size of the experimental snapshots (and not the
variability due to the width of the FFT peak, see below).

In Fig.~\ref{FFTexp} we present a couple of examples
(for $\omega/(2\pi)=140$~Hz and $\omega/(2\pi)=300$~Hz)
of the Faraday patterns observed in the experiments of
Ref.~\cite{engels}. Also in the figure, we depict the
FFT of the data (second row) as well as its $y$-integrated
counterpart (bottom row). As it is clear from the figure,
there is a dominant spatial frequency that can be measured
in order to extract the associated spacing of the Faraday
pattern. Nonetheless, it is important to note that the
dominant peak in the FFT spectrum has a width that
indicates an {\em interval} of spacings that make up
the original Faraday pattern. The width of this peak gives
an indication of spatial variability of the pattern spacing:
the wider the spectral peak the more spatial variability
there exists in the pattern spacing.
We have observed the same phenomenology when using
our 3D numerical data. A typical example of the FFT analysis
of our 3D numerical data is presented in Fig.~\ref{FFTnum}.
We have checked that our numerical data is able to reproduce
the behavior of the FFT analysis of the experimental data.
For a better comparison with experiments we added a 5\%
random noise to the numerically computed density so as to
emulate the noise in the experiment (see bottom
panel of Fig.~\ref{FFTnum}).
In order to extract the average spacing, $S_{\rm 3D}$, from
the 3D numerics we computed the position of the dominant peak
in the FFT (see pink square points in Fig.~\ref{fig_e1}).
As it is clear from the integrated FFT curves (see bottom
two panels in Fig.~\ref{FFTnum}), the Faraday pattern
contains a dominant peak with a respective finite width.
The width of the peak indicates the presence of a {\em range} of
spatial frequencies (instead of a single one) and thus
we can associate an error bar to the average spacing by
taking the width of the dominant peak as the variability
in spatial frequencies. This variability has been incorporated
in the average spacing of 3D numerical data in Fig.~\ref{fig_e1}
by means of the vertical error bars for the pink square points.
In the same spirit,
the actual error bar in the experiment should be slightly
larger to also account for variability of the spatial
frequency due to the width of the computed FFT peaks from
the experimental data.
Within this variability that, based on Figs.~\ref{FFTexp} and~\ref{FFTnum},
clearly exists both in the experimental and the numerical (3D)
data, we can conclude that our theoretical results are in very good
agreement with both their experimental and their numerical counterparts.

\section{Conclusions}

In this communication, we have revisited the topic of Faraday waves
and corresponding resonances in Bose-Einstein condensates. In particular,
we have focused on the experimentally relevant case where the transverse
confinement is periodically modulated in time. We have used the
non-polynomial Schr{\"o}dinger equation as a tool that permits to
present in an explicit form this transverse modulation in an effective
longitudinal equation describing the dynamics of the condensate
wavefunction. Then, a subsequent modulational stability analysis
permitted us to examine the stability of spatially uniform states
in this transversely driven, yet effectively one-dimensional setting.
This analysis, leading to a Mathieu equation, combined with
the well-established theory of the latter equation allowed us
to identify the dominant mode of the instability. This, in turn,
led us to extract an
explicit analytical formula that allows for this mode's wavenumber
(and hence its wavelength which is directly associated with the
pattern periodicity and hence is an experimentally observable quantity)
as a function of the driving frequency of the transverse confining
potential. Direct comparison of the fully analytical result with
the experimental observations confirmed the accuracy of our approach.
These analytical and experimental results were also corroborated
by numerical computations both within the framework of
the one-dimensional NPSE equation,
as well as for the case
of the fully 3D Gross-Pitaevskii equation. The similarities
of the two regarding the instability length scale and the differences
of the two in connection to the instability growth rate have been
accordingly highlighted. These computations have allowed us to
validate the quality of our theoretical approximations and
give a detailed comparison between theory, numerics
and experiment.

There are numerous interesting possibilities that this experiment
presents for future studies. One of them is to consider the
predominantly two-dimensional case of pancake-shaped condensates,
where, depending on the driving frequency, square or rhombic patterns
may form. In that setting too, the effective wave equations
of Ref.~\cite{npse} may allow to perform the modulational stability
analysis and obtain a quantitative handle on the dominant unstable
mode. On the other hand, modulations of all three spatial
dimensions of the confining potential would be of interest
in their own right. In the latter case, while reductions of the
type used herein would not be relevant, nevertheless the dynamics
may still be analytically describable upon appropriate assumptions,
such as, e.g., the quadratic phase assumption of Ref.~\cite{perez}, by
means of coupled, nonlinear ordinary differential equations
such as Eqs.~(12) of Ref.~\cite{perez}. Understanding these settings
in more quantitative detail, both analytically, numerically and
experimentally, is under current examination and will be reported
in future publications.

\medskip
\section*{Acknowledgments}
We are extremely thankful to Peter Engels for providing
numerous fruitful discussions, for
experimental data for Figs.~\ref{fig_e1} and~\ref{FFTexp},
and for carefully
reading this manuscript and suggesting numerous corrections
and additions.
AIN kindly acknowledges the help of Lisbeth Dilling, the librarian
of the Niels Bohr Institute, on the history of
Eq.~(\ref{critical_exponent}). PGK and RCG gratefully acknowledge
support from NSF-DMS (0505663, 0619492 and CAREER).
The authors acknowledge fruitful discussions
with Mogens H. Jensen, Mogens T. Levinsen and Henrik Smith.


\begin{thebibliography}{1}
\bibitem{cross} M.C. Cross and P.C. Hohenberg,
Rev. Mod. Phys. {\bf 65}, 851 (1993).

\bibitem{far} M. Faraday, Philos. Trans. R. Soc. London {\bf 121}, 299
(1831).

\bibitem{book1} C.J. Pethick and H. Smith,
{\it Bose-Einstein condensation in dilute gases}, Cambridge University
Press (Cambridge, 2002).


\bibitem{book2}  L.P. Pitaevskii and S. Stringari,
{\it Bose-Einstein Condensation}, Oxford University Press (Oxford, 2003).


\bibitem{perez} J.J. Garc{\'i}a-Ripoll, V.M. P{\'e}rez-Garc{\'i}a
and P. Torres, Phys. Rev. Lett. {\bf 83}, 1715 (1999).

\bibitem{dalfovo1} C. Tozzo, M. Kr{\"a}mer, and F. Dalfovo,  Phys. Rev. A
{\bf 72}, 023613 (2005).

\bibitem{dalfovo2} M. Kr{\"a}mer, C. Tozzo and F. Dalfovo,  Phys. Rev. A 71, 061602(R) (2005)

\bibitem{stoferle}
C. Schori, T. St{\"o}ferle, H. Moritz, M. K{\"o}hl, and T. Esslinger
Phys. Rev. Lett. {\bf 93}, 240402 (2005).

\bibitem{dalfovo3} M. Modugno, C. Tozzo, F. Dalfovo,
Phys. Rev. A {\bf 74}, 061601(R) (2006).

\bibitem{staliunas} K. Staliunas, S. Longhi and G.J. de Valc{\'a}rcel,
Phys. Rev. Lett. {\bf 89}, 210406 (2002).

\bibitem{feshbach} See e.g. S. Inouye {\it et al.},
Nature {\bf 392}, 151
(1998).

\bibitem{engels} P. Engels, C. Atherton and M.A. Hoefer,
Phys. Rev. Lett. {\bf 98}, 095301 (2007).


\bibitem{staliunas_second} K. Staliunas, S. Longhi and G.J. de Valc{\'a}rcel,
Phys. Rev. A {\bf 70}, 011601(R) (2004).

\bibitem{kagan}
Yu. Kagan and L. A. Manakova, Phys. Lett. A {\bf 361}, 401 (2007);
Yu. Kagan and L. A. Manakova, Phys. Rev. A {\bf 76}, 023601
(2007).

\bibitem{eng13} M. Fliesser, A. Csord{\'a}s, P. Sz{\'e}pfalusy and R. Graham,
Phys. Rev. A {\bf 56}, R2533 (1997).


\bibitem{eng14} S. Stringari, Phys. Rev. A {\bf 58}, 2385 (1998).

\bibitem{npse} L. Salasnich, A. Parola and L. Reatto,
Phys. Rev. A {\bf 65}, 043614 (2002).

\bibitem{book_of_McLachlan}N. W. McLachlan, \emph{Theory and application of Mathieu functions},
Oxford University Press, 1951.

\bibitem{mathematica}S. Wolfram, \emph{The Mathematica book}, Wolfram Media/Cambridge University
Press, 1999.
\bibitem{critical_exponent}
F{\'e}lix Tisserand, \emph{Trait{\'e} de m{\'e}canique
c{\'e}leste}, 1894, vol III (see the first chapter on how to
derive equation (\ref{critical_exponent})); G. W. Hill, \emph{Acta
Mathematica} \textbf{VIII}, 1886 (reprinted in \emph{Collected
Mathematical Works}, vol I, p. 255); E. L. Ince, \emph{Monthly
Notices of the Royal Astronomical Society} \textbf{LXXV}, 1915, p.
436; J. C. Adams, \emph{Collected Scientific Papers}, vol I, p.
186, vol II, pp. 65, 86; E. T. Whittaker, \emph{Proceedings of the
Edinburgh Mathematical Society} \textbf{XXXII}, 1914, p. 75.

\bibitem{cond_length}

The length of the condensate is computed assuming a Thomas-Fermi
density profile both radially and longitudinally. For the setup
described in Ref. \cite{engels} one obtains a length of 180
microns, i.e., $L=90\mu$m.

\bibitem{string1} S. Stringari,
Phys. Rev. Lett. {\bf 77}, 2360 (1996).

\bibitem{perez2} V. M. Perez-Garcia, H. Michinel, J. I. Cirac,
M. Lewenstein, and P. Zoller, Phys. Rev. Lett. {\bf 77}, 5320 (1996).

\end{thebibliography}
\end{document}